\documentclass[letterpaper,titlepage,11pt]{article}

\usepackage{jheppub}

\usepackage[T1]{fontenc}
\usepackage[utf8]{inputenc} 
\usepackage[turkish,english]{babel}

\usepackage{amsmath,amssymb,amsfonts}
\usepackage{esint}
\usepackage{empheq}
\usepackage{multirow}
\usepackage{enumerate}
\usepackage{needspace}

\usepackage{xcolor}
\usepackage{tikz}

\usepackage{textcomp}
\usepackage{bbm}

\usepackage[active]{srcltx}

\DeclareMathOperator{\Tr}{Tr}

\newcommand{\Lt}{{\tt L}}

\title{\boldmath\color{black}{Holonomies and Boundary Symmetries in Discrete BF Formulation of Jackiw--Teitelboim Gravity}}
\subheader{ITU Preprint,\,\\ \today}
\newcommand{\itu}{\dagger}

\author[\itu]{H.~T.~\"Ozer}
\emailAdd{ozert@itu.edu.tr}

\author[\itu]{Ayt\"ul~Filiz}
\emailAdd{aytulfiliz@itu.edu.tr}

\affiliation[\itu]{Istanbul Technical University,\,Faculty of Science and Letters,
\,Physics Department,\\34469 Maslak,\,Istanbul,Turkey.}
\abstract{ 
We develop a fully discrete and non-perturbative realization of two--dimensional Jackiw--Teitelboim (JT) gravity as an $\mathrm{sl}(2,\mathbb{R})$ BF theory. Using group--valued holonomies and Lie-algebra--valued dilatons, the bulk theory is shown to be purely topological, with all physical information encoded at the boundary. We analyze admissible discrete boundary conditions and derive the corresponding asymptotic symmetry algebras directly at the lattice level, including an affine Kac--Moody symmetry and its Brown--Henneaux reduction to a Virasoro algebra, together with the associated Virasoro--dilaton structure. A precise operator product expansion\,(OPE) dictionary is established by taking the controlled continuum limit of the discrete Poisson brackets. Beyond asymptotic symmetries, we provide an effective boundary description and a representation-theoretic quantization organized by monodromy sectors. Within this discrete framework, black hole entropy follows from gauge--invariant holonomy data and is expressed in terms of the dilaton Casimir, 
reproducing the Bekenstein--Hawking result within a framework that does not require an explicit Schwarzian action.
}
\begin{document}
\maketitle
\newpage

\section{Introduction}
\label{sec:intro}\vspace{0.5cm}
Two--dimensional JT gravity occupies a distinguished place among lower--dimensional models of gravity, as it provides a consistent and tractable setting for studying black hole
thermodynamics, holography, and quantum aspects of gravity
\cite{Jackiw1984,Teitelboim1983}. From a modern perspective, JT gravity
is most naturally described as a gauge theory, namely as a BF theory
with gauge group $\mathrm{SL}(2,\mathbb{R})$, where the gravitational
degrees of freedom are encoded in a Lie--algebra valued connection and a
dilaton field acting as a Lagrange multiplier
\cite{Ikeda1994,SchallerStrobl1994}. In this setting, the equations
of motion reduce to the flatness of the connection and the covariant
constancy of the dilaton, rendering the bulk theory topological from the
outset, a feature that underlies both continuum and higher--rank BF
extensions of JT gravity
\cite{Witten1991,Ozer:2025bpb}.

Within this gauge--theoretic perspective, asymptotic symmetries of JT gravity do not
arise as fundamental input but rather emerge as derived structures once appropriate
boundary conditions are imposed. In particular, the appearance of conformal symmetry
and the Virasoro algebra is intimately tied to specific choices of boundary conditions
and gauge fixings, such as the Brown--Henneaux or Drinfeld--Sokolov gauges~\cite{Brown:1986nw}.
This viewpoint highlights that asymptotic symmetries are not intrinsic to the bulk
theory itself, but are instead properties of the boundary phase space, as has been
explicitly demonstrated in continuum BF approaches to JT gravity~\cite{Grumiller:2017qao} 
and, more recently, in higher-rank generalizations~\cite{Ozer:2025bpb}.

In recent years, JT gravity has attracted renewed attention in the
context of AdS$_2$ holography, where boundary dynamics are often
described in terms of an effective action. Most prominently, the
Schwarzian theory has emerged as a universal low--energy description of
nearly-AdS$_2$ systems, capturing the explicit breaking of
reparametrization invariance down to $\mathrm{SL}(2,\mathbb{R})$
\cite{MaldacenaStanford2016,Jensen2016,EngelsoyMertensVerlinde2016}. While
this approach has proven extremely powerful, it is important to stress
that the Schwarzian action is not a fundamental ingredient of JT
gravity but rather an effective boundary description that arises after
specific gauge choices and continuum assumptions are imposed. This
interpretation is fully consistent with BF constructions of JT gravity
in the continuum, including higher--rank generalizations where boundary
symmetries and their breaking are derived directly from bulk gauge data
without the need to postulate an explicit boundary action
\cite{Grumiller:2017qao,Ozer:2025bpb}.

From this perspective, it is natural to ask whether the appearance of affine and Virasoro
structures in JT gravity is intrinsically tied to specific continuum choices,
such as boundary conditions, gauge fixings, and limiting procedures, or whether these
structures are already encoded more fundamentally at the level of gauge-invariant data.
The motivation for developing a discrete construction is therefore not to introduce new
asymptotic symmetry algebras, but to clarify why the familiar ones are unavoidable.
By describing the theory directly in terms of group--valued holonomies and monodromy data,
the discrete approach eliminates local bulk degrees of freedom from the outset and makes
manifest that affine and Virasoro symmetries arise as structural consequences of the boundary
phase space, rather than as assumptions or artifacts of continuum descriptions.

The gauge-theoretic description of JT gravity is not limited to
continuum approaches. On the contrary, it naturally suggests that
group-valued holonomies and conjugacy classes provide the most
fundamental and gauge-invariant variables through which the global
structure of the theory can be captured. From the BF perspective, 
such variables arise inevitably once the theory
is discretized, preserving the topological nature of the bulk while
making global and non-perturbative aspects manifest
\cite{Baez1996,Oriti2009,CattaneoFelder2000}. Related approaches
to discrete gravity, based on simplicial and geometric discretizations,
have been developed from a complementary perspective, emphasizing the
definition of curvature, symmetry principles, and black hole solutions
directly at the discrete level \cite{Chamseddine2021,Chamseddine2022,Chamseddine2023}.
In this sense, a lattice realization of JT gravity should not be viewed as an approximation to a
continuum theory, but rather as an intrinsic and faithful implementation of its underlying gauge
structure. From a broader topological perspective, holonomy-based discretizations are also known
to arise in three-dimensional Chern--Simons gravity, where group-valued variables naturally encode
flat connections~\cite{Dupuis:2017otn}. In the present work, however, our focus remains strictly
two-dimensional, aiming to derive the boundary phase space and its asymptotic symmetry structure
directly within a discrete BF description of JT gravity.

In this work, we investigate two--dimensional JT gravity
through a fully discrete BF realization. Our approach is based on
group-valued holonomies associated with lattice edges and
Lie-algebra valued dilaton variables assigned to vertices, preserving
the gauge-theoretic and topological character of the theory at every
stage, as naturally suggested by the BF description of two-dimensional
gravity \cite{Ikeda1994,SchallerStrobl1994,Witten1991}. By working
directly at the lattice level, we provide a
non-perturbative description in which all bulk degrees of freedom are
eliminated by flatness constraints, and the physical content is encoded
entirely in boundary holonomies and their conjugacy classes, in close
analogy with continuum BF approaches to JT gravity and their
higher-rank generalizations \cite{Ozer:2025bpb,Grumiller:2017qao},
reflecting the general insight that discretization in topological and
diffeomorphism-invariant theories should be viewed as a structural
feature rather than an approximation \cite{Dittrich2008}, a viewpoint that is also supported
by recent developments in discrete gravity where fundamental symmetries
are implemented directly at the lattice level \cite{ChamseddinePoincare2025}.
 
Within this discrete setting, we analyze how different choices of
boundary conditions lead to distinct asymptotic symmetry algebras at
the lattice level. In particular, we show how affine Kac--Moody
symmetries naturally arise from residual boundary gauge transformations
and how controlled reductions to Virasoro algebras can be implemented,
mirroring the structure found in continuum analyses based on BF
theories and asymptotic symmetry considerations
\cite{Brown:1986nw,Grumiller:2017qao,Ozer:2025bpb}. This
lattice-based realization provides a direct implementation of asymptotic
symmetry breaking in JT gravity and offers a transparent bridge between
bulk gauge data, boundary phase space, and the continuum limit.

In this work we construct JT gravity directly at the lattice
level, without invoking any continuum limit.
Throughout, discreteness should not be interpreted as a numerical approximation,
but rather as an exact rewriting in terms of group-valued holonomies and
Lie-algebra valued fields.
As in lattice gauge theory, local gauge covariance forbids the use of bare finite
differences between neighboring sites.
Instead, all discrete derivatives are understood in a gauge-covariant sense,
implemented via parallel transport by link variables.
This mechanism is standard in lattice gauge theory, beginning with Wilson’s original formulation of
lattice gauge theory \cite{Wilson1974}, appears systematically in the Hamiltonian approach of
Kogut \cite{Kogut1979}, and is made explicit in modern treatments under the name of
gauge-covariant (forward) differences \cite{Rothe2005}.

The discrete approach developed in this work also provides a natural
framework for addressing several aspects of JT gravity
that are traditionally discussed in the continuum. In particular, the
lattice description allows for a transparent treatment of boundary
charges, their Poisson structure, and the associated representation
theory, all of which are encoded directly in gauge-invariant holonomy
data, in line with the general properties of BF theories and their
boundary phase space descriptions
\cite{Witten1991,Grumiller:2017qao}. This perspective proves
especially useful when discussing the continuum limit, where the
discrete symmetry algebras reduce to their familiar continuum
counterparts in a controlled manner, as observed in continuum BF approaches
to JT gravity and its higher--rank generalizations
\cite{Ikeda1994,Ozer:2025bpb}.

Structures that in the continuum description are often introduced implicitly or recovered only 
after taking suitable limits can be defined directly and concretely within the discrete approach. 
In particular, gauge invariance, the flatness condition, and residual symmetries are expressed 
in terms of group-valued holonomies and lattice variables, without relying on fall--off conditions 
or functional analytic assumptions. This makes it manifest that asymptotic symmetries are 
localized on boundary vertices and links, and clarifies how affine and Virasoro-type structures 
emerge as continuum limits of well-defined discrete boundary algebras.

Although there exists an extensive body of work on two-dimensional JT gravity 
and its BF description, encompassing continuum approaches, boundary actions, and effective 
treatments, the present work intentionally restricts its use of references to a limited and 
representative subset. Our goal is not to provide a comprehensive survey of the JT literature,
 but rather to sharpen the focus on the structural role of discreteness and gauge--invariant 
 holonomy data, without diluting the discussion through an exhaustive enumeration of related studies. 
 Accordingly, we mainly refer to the original formulations of JT gravity and BF theory, together 
 with a small number of key works that establish the continuum boundary perspective, and concentrate 
 on developing the discrete construction and its implications in a self-contained manner.

The purpose of this work is to provide a fully discrete and non--perturbative description of 
JT gravity in which boundary dynamics, asymptotic symmetries, and black hole 
entropy emerge directly from gauge-invariant holonomy and monodromy data. Rather than postulating 
a boundary action or relying on continuum assumptions from the outset, we show that affine and
 Virasoro symmetries, their OPEs, and the entropy arise as structural 
 consequences of the discrete BF realization once appropriate boundary conditions are imposed. 
 In this sense, Schwarzian and Cardy descriptions should be regarded as effective continuum 
 languages capturing the same low-energy boundary dynamics that are already encoded in a more 
 fundamental discrete setting.

The central result of this work is the construction of a fully discrete,
non-perturbative BF realization of JT gravity in which the boundary phase
space and its associated structures arise directly from gauge-invariant
holonomy data. In this description, asymptotic symmetries and black hole
entropy are not introduced through boundary actions or continuum
assumptions, but emerge as structural consequences of the reduced
discrete phase space.

The paper is organized as follows. In Section~\ref{sec:discreteJT}, we describe
two-dimensional Jackiw--Teitelboim gravity as a discrete BF theory and introduce
the fundamental lattice variables together with the corresponding constraints.
Section~\ref{sec:ASA} analyzes asymptotic symmetries at the lattice level, including
both affine and Brown--Henneaux-type boundary conditions. In Section~\ref{sec:OPE},
we discuss the continuum limit and establish its connection with continuum BF
approaches and the associated OPEs. 
Section~\ref{sec:effective} develops the effective boundary description, including
the reduction to boundary data, quantization, and the emergence of entropy.
In Section~\ref{sec:entropy}, we analyze the entropy in terms of monodromy sectors
and the dilaton Casimir, and examine the consistency of the continuum limit with
the Cardy formula. Section~\ref{sec:holonomy} places the discrete construction in a
broader topological context by relating it to holonomy-based discretizations and
their dimensional reduction from three to two dimensions. 
Finally, Section~\ref{sec:conclusion} summarizes our results and outlines possible
extensions to higher-rank gauge groups and supersymmetric generalizations.
Appendix~\ref{app:appendices} provides the detailed Fourier-mode analysis of the lattice affine
Kac--Moody and Virasoro algebras, including their continuum limits and a comparison
with previous lattice constructions.

\section{\color{black}Discrete BF Formulation of JT Gravity}
\label{sec:discreteJT}

In this section we present a fully discrete realization of two-dimensional
JT gravity as a BF theory. Our construction follows the viewpoint that
discretization in gravity should be understood as a structural implementation
of the underlying gauge symmetries rather than as a mere approximation of a
continuum theory, as emphasized in diffeomorphism-invariant and topological
models of quantum gravity \cite{Dittrich2008}. Related developments in discrete
gravity, based on geometric and simplicial discretizations, further support
this perspective by demonstrating that fundamental symmetry principles can be
implemented directly at the lattice level \cite{Chamseddine2021}.

Within the BF description of JT gravity, this viewpoint finds a natural
realization. The continuum fields are replaced by group-valued holonomies
associated with lattice links and Lie-algebra valued dilaton variables assigned
to plaquettes, while flatness of the connection and covariant constancy of the
dilaton are implemented exactly as discrete constraints. The bulk theory thus
remains purely topological at the lattice level, and all nontrivial physical
information is encoded in boundary data (as in continuum BF treatments of JT
gravity, see e.g. \cite{Ikeda1994,SchallerStrobl1994,Witten1991}). This discrete
BF construction provides the starting point for the analysis of boundary
conditions and asymptotic symmetries carried out in the subsequent sections.

We consider a two-dimensional oriented manifold $\mathcal K$ with boundary,
whose topology is that of a disk. The manifold is discretized by a cellular
decomposition into vertices $v$, oriented links $\ell$, and plaquettes $f$.
No metric structure is assumed on the lattice, reflecting the topological
nature of the underlying BF theory.
\color{black}
The absence of any metric structure at the lattice level does not imply the absence of gravitational dynamics. Rather, the gravitational interpretation emerges after imposing appropriate boundary conditions and identifying the reduced boundary phase space, in complete analogy with continuum BF formulations of JT gravity. In this sense, the present construction should be understood not as a discretization of the metric formulation of JT gravity, but as a discrete realization of its underlying BF gauge structure, from which the gravitational interpretation arises at the boundary.
\color{black}

In the discrete gauge-theoretic description, the fundamental variables are
group-valued holonomies rather than local gauge fields. Accordingly, the
passage from the continuum description of JT gravity to a discrete BF
representation should be understood as a rewriting in terms of gauge-invariant
variables. In the continuum, JT gravity admits a BF description with
$\mathrm{SL}(2,\mathbb{R})$ gauge group, whose fundamental fields are a
Lie-algebra valued connection $\mathcal A$ and a dilaton multiplet
$\mathcal X$, and whose equations of motion enforce flatness of the connection
and covariant constancy of the dilaton
\cite{Jackiw1984,Teitelboim1983,Ikeda1994,SchallerStrobl1994,Witten1991}.
A consistent discretization must therefore preserve gauge covariance and the
topological character of the bulk \cite{Dittrich2008}.

In the discrete setting, the connection $\mathcal A$ is replaced by
group-valued holonomies $\mathcal U_\ell\in\mathrm{SL}(2,\mathbb{R})$
associated with oriented links $\ell$,
\begin{equation}
U_\ell
=
\mathcal{P}
\exp\!\left(
\int_\ell A_\mu(x)\, dx^\mu
\right),
\qquad
A = A_\mu(x)\, dx^\mu \in \mathfrak{sl}(2,\mathbb{R}) .
\end{equation}
where $A = A_\mu dx^\mu$ is an $\mathfrak{sl}(2,\mathbb{R})$-valued
connection one-form, so that its path-ordered exponential defines
a group-valued holonomy $U_\ell \in SL(2,\mathbb{R})$.
These holonomies transform covariantly under lattice gauge transformations.
 Gauge invariance forbids the use of bare finite differences between neighboring
sites, requiring discrete derivatives to be implemented via parallel
transport along links \cite{Wilson1974,Kogut1979,Rothe2005}. Ordered products
of link holonomies along oriented plaquettes are standard in non-abelian
lattice gauge theory, originating from Wilson’s formulation
\cite{Wilson1974} and systematized in the Hamiltonian approach of Kogut
\cite{Kogut1979} and in modern treatments \cite{Rothe2005}.

In the continuum limit, the flatness condition of BF theory becomes the exact
group--valued constraint $\mathcal W_f=\mathbbm{1}$ for all bulk plaquettes.

The dilaton multiplet is discretized as Lie-algebra valued variables
$\mathcal X_v$ assigned to vertices, transforming in the adjoint
representation. 
\color{black}
The discrete covariant derivative acting on the dilaton is defined along each oriented link $\ell$ as
\begin{equation}
(D\mathcal X)_\ell =\mathcal  X_{t(\ell)} -\mathcal U_\ell^{-1}\mathcal X_{s(\ell)}\mathcal U_\ell \,.
\end{equation}
This expression is the direct discrete analogue of the continuum covariant derivative
$D_\mu \mathcal X = \partial_\mu\mathcal X + [\mathcal A_\mu, \mathcal X]$,
where parallel transport is implemented by the link holonomies.
\color{black}
The continuum condition $D \mathcal X = 0$ is implemented at the discrete level by imposing
\begin{equation}
\mathcal X_{t(\ell)}=\mathcal U_\ell^{-1}\mathcal X_{s(\ell)}\mathcal U_\ell \,.
\end{equation}
This implies that the covariant discrete derivative vanishes. As a consequence,
all local bulk degrees of freedom are eliminated, and the gauge-invariant
physical content of the theory is encoded in boundary holonomies and their
conjugacy classes, in analogy with continuum BF descriptions and their
boundary phase space structures
\cite{Witten1991,Grumiller:2017qao,Ozer:2025bpb}.

Accordingly, to each oriented link $\ell$ we associate
\begin{equation}
\mathcal U_\ell \in \mathrm{SL}(2,\mathbb{R}),
\end{equation}
with orientation reversal corresponding to
\begin{equation}
\mathcal U_{\ell^{-1}} = \mathcal U_\ell^{-1}.
\end{equation}

Discrete curvature is encoded in the plaquette holonomy,
\begin{equation}
\mathcal W_f=\prod_{\ell\in\partial f}\mathcal U_\ell ,
\end{equation}
where the ordered product follows the orientation of $\partial f$. In the
continuum limit, $\mathcal W_f$ reduces to the exponential of the curvature
integrated over the plaquette, so that $\mathcal W_f=\mathbbm{1}$ corresponds
to vanishing curvature.

In JT gravity written as a BF theory, the dilaton acts as a Lagrange multiplier
enforcing flatness. In the discrete theory, we associate a Lie-algebra valued
dilaton variable
\begin{equation}
\mathcal X_f \in \mathfrak{sl}(2,\mathbb{R})
\end{equation}
to each plaquette (equivalently to the dual vertex).

The discrete BF action is defined as
\begin{equation}
\mathcal S_{\mathrm{BF}}^{\mathrm{disc}}
=
\sum_f \Tr\!\left(\mathcal X_f\,\log \mathcal W_f\right)
+\mathcal S_{\mathrm{boundary}},
\label{eq:discBFaction}
\end{equation}
where $\log\mathcal W_f$ maps the plaquette holonomy to the Lie algebra.

Variation with respect to $\mathcal X_f$ yields
\begin{equation}
\mathcal W_f = \mathbbm 1
\qquad \text{for all plaquettes } f ,
\end{equation}
which enforces local flatness, in direct analogy with
$\mathcal F=0$ in the continuum.

Variation with respect to $\mathcal U_\ell$ yields discrete covariant
constancy conditions for the dilaton. Together, flatness and covariant
constancy eliminate all local bulk degrees of freedom and confirm that the
discrete theory remains topological in the bulk.

Since $\mathcal K$ has a boundary, the bulk action
\eqref{eq:discBFaction} must be supplemented by boundary terms to ensure a
well-defined variational principle. The choice of boundary term determines
which boundary data are fixed.

The boundary is represented by a closed chain of oriented links
$n=0,\dots,N-1$, with boundary holonomies
\begin{equation}
\mathcal U_n \in \mathrm{SL}(2,\mathbb{R}).
\end{equation}
While bulk holonomies are constrained by flatness, boundary holonomies are
unconstrained and encode the physical phase space.

Different boundary terms correspond to different restrictions on boundary
variations. One may preserve an affine $\mathrm{SL}(2,\mathbb{R})$ symmetry,
leading to a discrete Kac--Moody algebra, or impose additional constraints
that reduce the symmetry to a discrete Virasoro algebra via a
Brown--Henneaux or Drinfeld--Sokolov-type reduction.

Thus, the discrete BF framework separates bulk and boundary dynamics: the
bulk enforces flatness and covariant constancy, while nontrivial degrees of
freedom and symmetry structures reside at the boundary. This separation will
be analyzed in detail in the next section.
\section{Asymptotic symmetries at the lattice level}
\label{sec:ASA}
In the discrete BF description developed above, asymptotic symmetries arise as
residual gauge transformations that preserve a chosen set of boundary conditions.
The crucial difference is that, in the discrete setting, this mechanism is realized
directly at the level of boundary holonomies and lattice variables, without reference
to fall-off conditions or continuum gauge fixings.

In this section we analyze the asymptotic symmetry structure of discrete JT gravity arising 
from residual boundary gauge transformations that preserve the chosen lattice boundary 
conditions as in continuum BF analyses of JT gravity, see e.g.
\cite{Witten1991,Grumiller:2017qao,Ozer:2025bpb}.
In the present framework, however, this analysis is performed entirely at the lattice level. Bulk gauge
transformations act on the discrete variables $(\mathcal U_l,\mathcal X_f)$ in the standard
manner and are rendered pure gauge by the flatness constraints imposed by the
BF action. In the presence of a boundary, gauge transformations that act
trivially in the bulk but nontrivially on boundary vertices survive as physical
symmetries. As a consequence, all nontrivial gauge and holonomy data become
localized on the boundary links. The resulting residual gauge transformations
form an effective loop group of $\mathrm{SL}(2,\mathbb{R})$ at the boundary,
leading to a nontrivial algebra of boundary charges with an underlying affine
structure. This discrete realization provides a direct framework for analyzing boundary conditions, 
symmetry breaking patterns, and the resulting boundary dynamics, which will be explored 
in the following subsections.

\subsection{Affine boundary conditions}
\label{subsec:affine-KM}
We begin by imposing affine boundary conditions on the discrete boundary holonomies.
These conditions represent the least restrictive choice compatible with a well-defined
variational principle and preserve the full residual gauge symmetry at the boundary.
At the lattice level, bulk gauge transformations are rendered pure gauge by the flatness
constraints, while gauge transformations that act nontrivially on boundary links survive
as physical symmetries. As a consequence, the boundary holonomies furnish an effective
loop group of $\mathrm{SL}(2,\mathbb{R})$, and the associated boundary charges organize
themselves into an affine Kac--Moody algebra. This structure arises directly from the
discrete BF theory and does not rely on continuum gauge fixings or fall--off
conditions.

The boundary of the manifold is discretized as a circle parametrized by sites
$n=0,\dots,N-1$, with lattice spacing $\Delta\tau=2\pi/N$. Boundary links connect
neighboring sites $(n,n+1)$ and are associated with holonomies
\begin{equation}
\mathcal U_n = \mathcal{P}\exp\!\left(\int_{\tau_n}^{\tau_{n+1}} a_\tau(\tau)\,d\tau\right),
\qquad
\tau_n = n\Delta\tau .
\end{equation}
In the discrete theory, the $\mathcal U_n$ are taken as the fundamental boundary
variables.

To make contact with the Lie-algebraic description of boundary symmetries, it is
convenient to introduce the log--link variables
\begin{equation}
\mathcal U_n = \exp(\Delta\tau\, a_n),
\qquad
a_n \in \mathfrak{sl}(2,\mathbb R).
\end{equation}
In the continuum limit $\Delta\tau\to0$, $a_n$ reduces to the boundary component
$a_\tau(\tau)$ of the gauge connection.

The affine boundary condition is imposed by restricting $a_n$ to take the form
\begin{equation}
a_n = \alpha^i\,\mathcal L^{\,i}_{n}\,\Lt_i ,
\label{eq:affineBC}
\end{equation}
where $\Lt_i$ ($i=-1,0,+1$) are generators of $\mathfrak{sl}(2,\mathbb{R})$,
$\alpha^i$ are fixed constants specifying the boundary condition, and
$\mathcal L^i_n$ are dynamical boundary fields.
This discrete affine boundary condition mirrors the structure encountered in
continuum BF analyses of JT gravity and its higher-rank
generalizations, where affine boundary symmetries arise from residual gauge
transformations \cite{Grumiller:2017qao,Ozer:2025bpb,Witten1991}.
At this stage no constraint is
imposed on the $\mathcal L^i_n$. In this setup, gauge transformations that 
preserve the affine boundary condition act nontrivially on the boundary 
degrees of freedom and, in particular, on the boundary holonomies. For
the holonomy,
\begin{equation}
\mathcal U_n \;\longrightarrow\; g_n^{-1}\mathcal U_n g_{n+1},
\qquad
 g_n\in \mathrm{SL}(2,\mathbb{R}),
\end{equation}
with infinitesimal parametrization
\begin{equation}
g_n=\exp(\lambda_n),
\qquad
\lambda_n\in\mathfrak{sl}(2,\mathbb{R}).
\end{equation}

Using $\mathcal U_n=e^{\Delta\tau a_n}$, the transformation reads
\begin{equation}
\mathcal U_n \;\longrightarrow\; e^{-\lambda_n}e^{\Delta\tau a_n}e^{\lambda_{n+1}}.
\end{equation}
Applying the Baker--Campbell--Hausdorff formula to first order,
\begin{equation}
e^{-\lambda_n}e^{\Delta\tau a_n}e^{\lambda_{n+1}}
=
\exp\!\Big(
\Delta\tau a_n + (\lambda_{n+1}-\lambda_n)
+ \Delta\tau[a_n,\lambda_n]
+ \mathcal O(\lambda^2)
\Big).
\end{equation}
Taking the logarithm yields
\begin{equation}
\delta a_n
=
\frac{\lambda_{n+1}-\lambda_n}{\Delta\tau}
+
[a_n,\lambda_n].
\label{eq:disc-gauge-an}
\end{equation}
This identifies the gauge-covariant discrete derivative,
\begin{equation}
\nabla\lambda_n
=
\frac{\lambda_{n+1}-\lambda_n}{\Delta\tau}
\label{eq:forward-derivative},
\end{equation}
so that \eqref{eq:disc-gauge-an} becomes
\begin{equation}
\delta a_n=\nabla\lambda_n+[a_n,\lambda_n].
\label{eq:disc-gauge-an2}
\end{equation}

This is the discrete counterpart of the continuous affine transformation and,
in particular, the exact discrete analogue of the continuum relation
$\delta a_\tau = \partial_\tau \lambda + [a_\tau,\lambda]$.
Accordingly, affine boundary conditions are preserved by gauge parameters of the form
\begin{equation}
\lambda_n = \epsilon^i_n\,\Lt_i ,
\end{equation}
with arbitrary functions $\epsilon^i_n$ defined on the boundary lattice. Substituting
\eqref{eq:affineBC} into \eqref{eq:disc-gauge-an2} and projecting onto the
$\mathfrak{sl}(2,\mathbb R)$ basis yields the transformation laws
\begin{equation}
\delta \mathcal L^{\,i}_{n}
=
\frac{1}{\alpha^i}\,\nabla \epsilon^i_n
+
\sum_{j+k=i}(j-k)\,\frac{\alpha^j}{\alpha^i}\,
\mathcal L^{\,j}_{n}\,\epsilon^k_n .
\label{eq:affineTrafo}
\end{equation}

We now evaluate this expression explicitly for $i=+1,0,-1$.

Collecting the results, the discrete affine transformations take the form
\begin{align}
\delta\mathcal L^{\pm1}_{n}
&=
\frac{1}{\alpha^{\pm1}}\nabla\epsilon_n^{\pm1}
\pm \mathcal L^{\pm1}_{n}\epsilon_n^0
\mp \frac{\alpha^{0}}{\alpha^{\pm1}}\mathcal L^{0}_{n}\epsilon_n^{\pm1},
\label{eq:disc-affine-pm}\\[0.3em]
\delta\mathcal L^{0}_{n}
&=
\frac{1}{\alpha^{0}}\nabla\epsilon_n^{0}
+
\frac{2\alpha^{+1}}{\alpha^{0}}\mathcal L^{+1}_{n}\epsilon_n^{-1}
-
\frac{2\alpha^{-1}}{\alpha^{0}}\mathcal L^{-1}_{n}\epsilon_n^{+1}.
\label{eq:disc-affine-0}
\end{align}

These equations display the discrete $\mathfrak{sl}(2,\mathbb{R})$ affine structure
in a manifest three-component block form and act nontrivially on the boundary data,
thereby generating genuine asymptotic symmetries.

The generators of these transformations are obtained from the discrete BF
symplectic structure. Requiring that the charge $Q[\epsilon]$ generates the
transformation \eqref{eq:affineTrafo} via the Poisson bracket uniquely determines
its variation as
\begin{equation}
\delta Q[\epsilon]
=
\frac{k}{2\pi}\sum_n \Delta\tau\,
\Tr(\lambda_n\,\delta a_n).
\label{eq:deltaQ_connection}
\end{equation}
\color{black}
After imposing the bulk constraints, the reduced boundary phase space is parametrized by holonomy (or equivalently connection) variables. The canonical generators of residual gauge transformations are therefore naturally expressed in terms of variations of the connection, leading to eq.~\eqref{eq:deltaQ_connection}. This is consistent with continuum BF formulations, where the canonical charges admit both connection and dilaton contributions~\cite{Ozer:2025bpb}. 

An analogous expression can be written for the dilaton sector, $\delta Q[\epsilon] \sim \Tr(\lambda_n\,\delta\mathcal X_n)$, although the present formulation is naturally expressed in terms of connection/holonomy variables.
\color{black}

Assuming integrability, this yields the boundary charge
\begin{equation}
Q[\epsilon]
=
\frac{k}{2\pi}\sum_n \Delta\tau\,
\epsilon^i_n\,\alpha^i\,\mathcal L^{\,i}_{n}.
\end{equation}
\color{black}
with $\alpha^i = \frac{1}{k}; \{i=-1,0,+1\}$. The topological nature of the BF theory implies that the boundary charges are conserved under the discrete time evolution.

Using the defining relation
\begin{equation}
\delta_{\epsilon_2} a_n
=
\{a_n,Q[\epsilon_2]\},
\end{equation}
one finds that the boundary fields $\mathcal L^{\,i}_{n}$ satisfy the Poisson
brackets
\begin{equation}
\frac{\Delta\tau}{2\pi}\{\mathcal L^{\,i}_{n},\mathcal L^{\,j}_{m}\}
=
 (i-j)\,\mathcal L^{\,i+j}_{n}\,\delta_{n,m}
+
 k\,\eta^{ij}\,\nabla\delta_{n,m},
\label{eq:discKM}
\end{equation}

where $\eta_{ij}=\Tr(\Lt_i\Lt_j)$ is the invariant bilinear form on
$\mathfrak{sl}(2,\mathbb R)$. 
Equation~\eqref{eq:discKM} gives the lattice
current algebra in the site basis. 
The indices $i,j\in\{-1,0,+1\}$  are internal
$\mathfrak{sl}(2,\mathbb R)$ labels, whereas $n,m$ label boundary lattice sites;
they are not affine mode numbers.  Consequently, the conventional level-addition
rule is not expected to be visible in this basis.
The corresponding Fourier-mode algebra,
together with its continuum limit leading to the affine Kac--Moody structure,
is derived in Appendix~\ref{app:lattice-KM}. At this stage, the residual boundary 
symmetry is encoded entirely in the lattice current algebra.

\color{black}

While this affine structure represents the most general
residual gauge symmetry compatible with the discrete variational principle,
it can be consistently reduced by imposing additional boundary constraints.
Such reductions lead to a discrete Virasoro algebra, which will be analyzed
in the next subsection.
In the limit $\Delta\tau \to 0$, the discrete derivative and Kronecker delta
combine to reproduce the standard affine Kac--Moody algebra of the continuum
theory (see e.g. the continuum affine current algebra discussion in
\cite{Witten1991,Ozer:2025bpb}).
Thus, the discrete framework provides a faithful lattice realization of the classical 
affine boundary symmetry of JT gravity. In the BF description of JT gravity, 
the dilaton field is Lie-algebra valued and transforms in the adjoint representation. 
On the discrete boundary we introduce
\begin{equation}
\mathcal X_n = \mathcal X_n^i \Lt_i \in \mathfrak{sl}(2,\mathbb{R}),
\end{equation}
in complete analogy with the continuous dilaton field
$\mathcal X=\mathcal X^i(t)\Lt_i$.

Under an infinitesimal gauge transformation generated by
$\lambda_n=\epsilon_n^j \Lt_j$, the discrete dilaton transforms as
\begin{equation}
\delta_\epsilon\mathcal  X_n = [\mathcal X_n,\lambda_n].
\end{equation}
Unlike the connection $a_n$, no discrete derivative appears, since $\mathcal X_n$
is a site variable.

Using
\begin{equation}
[\mathcal X_n,\lambda_n]
=
\mathcal X_n^k\epsilon_n^j[\Lt_k,\Lt_j]
=
\mathcal X_n^k\epsilon_n^j(k-j)\Lt_{k+j},
\end{equation}
the component-wise variation is
\begin{equation}
\delta_\epsilon \mathcal X^{\,i}_{n}
=
\sum_{k+j=i}(k-j)\,\mathcal X^{\,k}_{n}\,\epsilon_n^{\,j}.
\label{eq:disc-dilaton-components}
\end{equation}

Evaluating \eqref{eq:disc-dilaton-components} for $i=+1,0,-1$ yields:
\begin{align}
\delta_\epsilon\mathcal  X^{\pm1}_{n}
&=
\pm \mathcal X^{\pm1}_{n}\,\epsilon_n^{0}
\mp \mathcal X^{0}_{n}\,\epsilon_n^{\pm1},
\label{eq:disc-dilaton-plus}
\\[0.3em]
\delta_\epsilon \mathcal X^{0}_{n}
&=
2\mathcal X^{+1}_{n}\,\epsilon_n^{-1}
-
2\mathcal X^{-1}_{n}\,\epsilon_n^{+1},
\label{eq:disc-dilaton-minus}
\end{align}
These relations hold pointwise for each lattice site $n$.

Several observations are immediate: the discrete dilaton transforms purely
algebraically and does not involve the discrete derivative $\nabla$, the
transformation laws coincide exactly with the continuous adjoint action upon
the replacement $t \to n$, and the dilaton therefore acts as a coadjoint element
stabilizing the discrete gauge symmetry, as in the continuous JT/BF theory.

Since the boundary charge is linear in $a_n$, the transformation laws
\eqref{eq:disc-dilaton-plus}--\eqref{eq:disc-dilaton-minus}
imply that the mixed Poisson brackets
$\{\mathcal L^{\,i}_{n},\,\mathcal X^{\,j}_{m}\}$
are fixed uniquely and take the form
\begin{equation}
\{\mathcal L^{\,i}_{n},\mathcal X^{\,j}_{m}\}
=
(2i+j)\,\mathcal X^{\,i+j}_{n}\,\delta_{n,m}
\label{eq:dilaton-weight-minus-one}.
\end{equation}
\color{black}
The mode-addition structure extends directly to the mixed
current--dilaton sector. Introducing the Fourier modes
\begin{equation}
\mathcal L_r^{\,i}
=
\frac{1}{N}
\sum_{n=0}^{N-1}L_n^{\,i}e^{-ir\tau_n},
\qquad
\mathcal X_s^{\,j}
=
\frac{1}{N}
\sum_{n=0}^{N-1}X_n^{\,j}e^{-is\tau_n},
\end{equation}
the site-space bracket ~\eqref{eq:dilaton-weight-minus-one} gives
\begin{align}
\left\{\mathcal L_r^{\,i},\mathcal X_s^{\,j}\right\}
&=
\frac{1}{N^2}
\sum_{n,m}
e^{-ir\tau_n-is\tau_m}
\left\{L_n^{\,i},X_m^{\,j}\right\}
\nonumber\\
&=
\frac{1}{N}
(2i+j)
\sum_n X_n^{\,i+j}e^{-i(r+s)\tau_n}
\nonumber\\
&=
(2i+j)X_{r+s}^{\,i+j}.
\end{align}
Therefore, one obtains the exact finite-lattice result
\begin{equation}
\left\{\mathcal L_r^{\,i},\mathcal X_s^{\,j}\right\}
=
(2i+j)\mathcal X_{r+s}^{\,i+j}.
\label{eq:mixed-affine-dilaton-Fourier}
\end{equation}
Thus, the mixed current--dilaton sector exhibits the exact mode-addition
rule already at finite lattice spacing, with no derivative-dependent lattice
correction and no central contribution.
\color{black}
\subsection{Conformal boundary conditions}
\label{subsec:BH-discrete}
Here we implement a discrete analogue of Brown--Henneaux boundary conditions,
which consistently reduces the affine Kac--Moody boundary symmetry identified
in Section~\ref{subsec:affine-KM} to a Virasoro algebra. From the gauge--theoretic perspective
developed above, this reduction is not imposed as an independent assumption,
but arises as a restriction of the affine boundary phase space compatible with
the discrete variational principle. In direct analogy with the Brown--Henneaux
and Drinfeld--Sokolov reductions in the continuum
~\cite{Brown:1986nw,Grumiller:2017qao,Ozer:2025bpb}, the role of these
boundary conditions is to eliminate redundant boundary degrees of freedom such
that the residual symmetry algebra is reduced from an affine Kac--Moody algebra
to a single copy of the Virasoro algebra.

Starting from the affine parametrization
\begin{equation}
a_n=\alpha^i \mathcal{L}^i_n L_i ,
\end{equation}
we impose the discrete Drinfeld--Sokolov gauge by fixing the highest-weight component and eliminating the $L_0$ component,
\begin{equation}
\alpha^{+1}\mathcal{L}^{+1}_n=1,\qquad
\alpha^{0}\mathcal{L}^{0}_n=0,\qquad
\alpha^{-1}=\alpha,\qquad
\mathcal{L}^{-1}_n\equiv \mathcal{L}_n .
\end{equation}
The boundary connection is then reduced to
\begin{equation}
a_n=L_{+1}+\alpha \mathcal{L}_n L_{-1}
\label{eq:BH-connection-alpha},
\end{equation}
where $\mathcal{L}_n$ is the single dynamical boundary field.
\color{black}

A general discrete gauge transformation
\begin{equation}
\delta a_n=\nabla\lambda_n+[a_n,\lambda_n]
\end{equation}
preserves the form \eqref{eq:BH-connection-alpha} provided the gauge parameter
\begin{equation}
\lambda_n=\epsilon_n^{+1}\Lt_{+1}
+\epsilon_n^{0}\Lt_0
+\epsilon_n^{-1}\Lt_{-1}
\label{eq:BH-connectionn}
\end{equation}
satisfies the constraints
\begin{align}
\epsilon_n^{0}
&=
-\,\nabla\epsilon_n^{+1},
\\[0.4em]
\epsilon_n^{-1}
&=
\alpha\,
\mathcal L_n\,\epsilon_n^{+1}
+
\frac{1}{2}\,\nabla^2\epsilon_n^{+1}.
\end{align}
All residual gauge freedom is therefore parametrized by a single function
\begin{equation}
\epsilon_n\equiv\epsilon_n^{+1}.
\end{equation}
Using the above relations, the variation of the remaining boundary field
$\mathcal L_n$ takes the form
\begin{equation}
\delta_\epsilon\mathcal L_n
=
\epsilon_n\,\nabla\mathcal L_n
+
2\mathcal L_n\,\nabla\epsilon_n
+
\frac{2}{\alpha}\,\nabla^3\epsilon_n.
\label{eq:disc-Virasoro-var-alpha}
\end{equation}
This is the discrete Virasoro transformation with central term proportional to
$\alpha$.
\color{black}
Equation~\eqref{eq:disc-Virasoro-var-alpha} gives the lattice Drinfeld--Sokolov
transformation in the site basis. 
The index $n$ denotes boundary lattice sites rather than Virasoro mode numbers. Consequently, 
the conventional level-addition rule is not expected to be visible in this basis.
Its Fourier-mode realization and the corresponding continuum Virasoro structure are derived 
in Appendix~\ref{app:lattice-Virasoro}.
\color{black}

In the BF description, the dilaton is a Lie-algebra valued field transforming in
the adjoint representation. On the discrete boundary we introduce
\begin{equation}
\mathcal X_n = \mathcal X_{n}^{+1} \Lt_{+1} + \mathcal X_{n}^{0} \Lt_{0} +\mathcal  X_{n}^{-1} \Lt_{-1}.
\end{equation}
Under an infinitesimal gauge transformation generated by
$\lambda_n$, the dilaton transforms as
\begin{equation}
\delta_\epsilon \mathcal X_n = [\mathcal X_n,\lambda_n].
\end{equation}

Compatibility with the Drinfeld--Sokolov gauge
\eqref{eq:BH-connectionn} requires fixing the highest-weight components of the
dilaton. We impose
\begin{equation}
\mathcal X_{n}^{+1} = \text{const.},
\qquad
\mathcal X_{n}^{0} = 0,
\end{equation}
so that the only dynamical dilaton component is $\mathcal X_{n}^{-1}$.

Using the constrained gauge parameter
\begin{align}
\epsilon_n^{0} &= -\,\nabla \epsilon_n,\\
\epsilon_n^{-1} &= \alpha\,\mathcal L_n\,\epsilon_n
+\frac{1}{2}\,\nabla^{2}\epsilon_n,
\end{align}
with $\epsilon_n\equiv\epsilon_n^{+1}$, the variation of the remaining dilaton
component becomes
\begin{equation}
\delta_\epsilon \mathcal X_{n}^{-1}
=
\epsilon_n\,\nabla \mathcal X_{n}^{-1}
-
\mathcal X_{n}^{-1}\,\nabla \epsilon_n .
\label{eq:disc-dilaton-BH}
\end{equation}

Equation~\eqref{eq:disc-dilaton-BH} shows that $\mathcal X_{n}^{-1}$ transforms as a
conformal field of weight $-1$ under the discrete Virasoro symmetry generated by
$\mathcal L_n$. This precisely mirrors the role of the dilaton in the continuous
Drinfeld--Sokolov reduction of JT gravity.

In Drinfeld--Sokolov gauge,
\begin{equation}
a_n = \Lt_{+1} + \alpha\,\mathcal L_n \Lt_{-1},
\qquad
\mathcal X_n \equiv \mathcal X_{n}^{-1},
\end{equation}
the residual symmetry is generated by a single lattice function $\epsilon_n$.
The dilaton transforms as a conformal field of weight $-1$,
\begin{equation}
\delta_\epsilon \mathcal X_n
=
\epsilon_n\,\nabla\mathcal X_n
-
\mathcal X_n\,\nabla\epsilon_n .
\label{eq:disc-dilaton-var}
\end{equation}
We define the boundary charge associated with $\epsilon_n$ as
\begin{equation}
Q[\epsilon]
=
\frac{k}{2\pi}\sum_m \Delta\tau\,\epsilon_m\,\mathcal L_m .
\label{eq:disc-Virasoro-charge}
\end{equation}
Requiring that the charge generates the transformation,
\begin{equation}
\delta_\epsilon\mathcal X_n
=
\{\mathcal X_n,Q[\epsilon]\},
\label{eq:disc-mixed-gen}
\end{equation}
uniquely fixes the mixed Poisson bracket. Writing
\begin{equation}
\{\mathcal X_n,\mathcal L_m\}
=
A\,\mathcal X_n\,\nabla\delta_{n,m}
+
B\,(\nabla\mathcal X_n)\,\delta_{n,m},
\end{equation}
and inserting into \eqref{eq:disc-mixed-gen} yields
\begin{equation}
\{\mathcal X_n,\mathcal L_m\}
=
\frac{2\pi}{k}\Big(
\mathcal X_n\,\nabla\delta_{n,m}
-
(\nabla\mathcal X_n)\,\delta_{n,m}
\Big).
\label{eq:disc-mixed-bracket}
\end{equation}
Equivalently, using antisymmetry,
\begin{equation}
\{\mathcal L_n,\mathcal X_m\}
=
\frac{2\pi}{k}\Big(
(\nabla\mathcal X_m)\,\delta_{n,m}
-
\mathcal X_m\,\nabla\delta_{n,m}
\Big).
\label{eq:disc-mixed-bracket-LX}
\end{equation}
This bracket reproduces the weight $-1$ transformation
\eqref{eq:disc-dilaton-var} and reduces in the continuum limit to the standard
Virasoro action on a primary field of weight $-1$. 

\color{black}
The Fourier-mode structure of the Virasoro--dilaton sector can likewise be
displayed directly in the forward representation. Introducing
\begin{equation}
\mathcal L_s
=
-\frac{i}{N}
\sum_{n=0}^{N-1}L_n e^{-is\tau_n},
\qquad
\mathcal X_r
=
\frac{1}{N}
\sum_{n=0}^{N-1}X_n e^{-ir\tau_n},
\end{equation}
and using the forward lattice momentum
\begin{equation}
\omega_p^{(+)}
=
\frac{e^{ip\Delta\tau}-1}{\Delta\tau},
\qquad
\nabla e^{ip\tau_n}
=
\omega_p^{(+)}e^{ip\tau_n},
\end{equation}
the Fourier transform of Eq.~\eqref{eq:disc-mixed-bracket-LX} gives
\begin{equation}
\left\{\mathcal L_s,\mathcal X_r\right\}
=
-i\left[
\omega_{r+s}^{(+)}
-
\omega_{-s}^{(+)}
\right]X_{r+s}.
\label{eq:Vir-dilaton-Fourier}
\end{equation}
The mode-addition structure is therefore already explicit at finite lattice
spacing. Since
\begin{equation}
\omega_p^{(+)}
=
ip+\mathcal{O}(\Delta\tau),
\end{equation}
the continuum limit becomes
\begin{equation}
\left\{L_s,X_r\right\}
=
(r+2s)X_{r+s}
+
\mathcal{O}(\Delta\tau).
\label{eq:Vir-dilaton-Fourier-cont}
\end{equation}

This is the standard Virasoro mode action on a
field of conformal weight $h=-1$. This completes the discrete
Virasoro--dilaton sector and its connection with the corresponding
continuum mode representation.
\color{black}
\section{OPE dictionary and continuum limit}
\label{sec:OPE}
We now establish the OPE dictionary by taking a controlled continuum limit of the discrete Poisson brackets. 
In this limit, Kronecker deltas and discrete derivatives combine into their continuum counterparts, 
allowing the boundary charge algebra derived at the lattice level to be rewritten in standard CFT language.

We consider the continuum limit defined by
\begin{equation}
\Delta\tau \rightarrow 0,
\qquad
N\Delta\tau = 2\pi,
\end{equation}
and identify lattice sites with points on the boundary circle,
\begin{equation}
\tau_n = n\Delta\tau .
\end{equation}
Discrete boundary fields are mapped to smooth functions according to
\begin{equation}
\mathcal L_n \;\longrightarrow\; \mathcal L(\tau),
\qquad
\mathcal X_n \;\longrightarrow\;\mathcal X(\tau),
\end{equation}
while discrete derivatives and Kronecker deltas are related to their continuum
counterparts via
\begin{equation}
\nabla f_n \;\longrightarrow\; \partial_\tau f(\tau),
\qquad
\Delta\tau\,\delta_{n,m} \;\longrightarrow\; \delta(\tau-\tau').
\label{eq:disc-cont-map}
\end{equation}
\color{black}
To connect the continuum fields with the Fourier-mode description developed
in Section~\ref{sec:ASA} and Appendices~\ref{app:lattice-KM}--\ref{app:lattice-Virasoro},
we use the standard mode--field correspondence.
With $z=e^{i\tau}$, the affine current, stress tensor, and dilaton fields are
associated with the Fourier modes $\mathcal L_r^{\,i}$, (or
$L_r$ after the Drinfeld--Sokolov reduction) and $\mathcal X_r^{\,i}$ (or
$X_r$ after the Drinfeld--Sokolov reduction), respectively. In particular,
\begin{equation}
\mathfrak{J}^{i}(z)
=
\sum_{r\in\mathbb{Z}}\mathcal L_r^{\,i}z^{-r-1},
\qquad
\mathcal{L}(z)
=
\sum_{r\in\mathbb{Z}}L_r z^{-r-2},
\label{eq:mode-field-dictionary}
\end{equation}
while the corresponding dilaton modes assemble into the continuum fields $\mathcal{X}^{i}(z)$ and $\mathcal{X}(z)$,

\begin{equation}
\mathcal{X}^{i}(z)
=
\sum_{r\in\mathbb{Z}}\mathcal X_r^{\,i}z^{-r-1},
\qquad
\mathcal{X}(z)
=
\sum_{r\in\mathbb{Z}}X_r z^{-r+1}.
\label{eq:mode-dilaton-field-dictionary}
\end{equation}
The mode algebras derived in Section~\ref{sec:ASA} 
and Appendices~\ref{app:lattice-KM}--\ref{app:lattice-Virasoro} are therefore 
related to the OPEs below by the standard mode--OPE correspondence.
\color{black}
\subsection{Affine OPEs}
\label{subsec:AffineOPE}

We now turn to the OPE associated with the affine boundary
symmetry algebra derived above. Translating the discrete Poisson brackets into the
OPE language, one obtains the classical affine Kac--Moody structure for the boundary
currents. This coincides with the standard classical affine Kac--Moody OPE
(see e.g. the standard affine current algebra in continuum BF and CFT approaches
\cite{Witten1991,Ozer:2025bpb}).
We begin with the affine boundary conditions discussed in
Section~\ref{subsec:affine-KM}. In the continuum limit, the discrete Poisson
brackets \eqref{eq:discKM} translate into the OPE
\begin{equation}
\mathfrak J^i(\tau_1)\,\mathfrak J^j(\tau_2)
\sim
\frac{k\,\eta^{ij}}{\tau_{12}^{2}}
+
\frac{f^{ij}{}_k\,\mathfrak J^k(\tau_2)}{\tau_{12}},
\qquad
\tau_{12}=\tau_1-\tau_2,
\label{eq:KM-OPE}
\end{equation}
where $\mathfrak J^i(\tau)$ are the continuum affine currents associated with
$\mathcal L_i(\tau)$, $\eta^{ij}$ is the invariant bilinear form on
$\mathfrak{sl}(2,\mathbb R)$, and $f^{ij}{}_k$ are the structure constants.

The adjoint transformation of the dilaton under affine symmetry obtained in the
discrete theory leads to the mixed OPE
\begin{equation}
\mathfrak J^i(\tau_1)\,\mathcal X^j(\tau_2)
\sim
\frac{f^{ij}{}_k\,\mathcal X^k(\tau_2)}{\tau_{12}},
\label{eq:KM-dil-OPE}
\end{equation}
which reflects the Lie-algebraic action of the affine symmetry on the boundary
dilaton field.
(see e.g. the standard current--adjoint-field OPE structure in continuum BF/CFT approaches \cite{Witten1991,Ozer:2025bpb}).
\subsection{Virasoro OPEs}
\label{subsec:VirasoroOPE}
We now turn to the Brown--Henneaux boundary conditions analyzed in
Section~\ref{subsec:BH-discrete}. In the continuum limit, the remaining boundary field
$\mathcal L(\tau)$ is naturally identified with the chiral stress tensor.
The discrete Virasoro algebra \eqref{eq:disc-Virasoro-var-alpha} then translates into the
standard Virasoro OPE

\begin{equation}
\mathcal L(\tau_1)\,\mathcal L(\tau_2)
\sim
\frac{\frac{c}{2}}{\tau_{12}^4}
+
\frac{2\,\mathcal L(\tau_2)}{\tau_{12}^2}
+
\frac{\partial \mathcal L(\tau_2)}{\tau_{12}},
\label{eq:TT-OPE}
\end{equation}
with central charge
\begin{equation}
c = 6k
\label{eq: central-charge}.
\end{equation}
(see e.g. the Virasoro OPE arising from Brown--Henneaux boundary conditions and
Drinfeld--Sokolov reduction in the continuum \cite{Brown:1986nw,Grumiller:2017qao}).

Finally, the mixed Poisson bracket \eqref{eq:disc-mixed-bracket-LX} yields the
Virasoro--dilaton OPE
\begin{equation}
\mathcal L(\tau_1)\,\mathcal X(\tau_2)
\sim
-\frac{\mathcal X(\tau_2)}{\tau_{12}^{2}}
+
\frac{\partial \mathcal X(\tau_2)}{\tau_{12}},
\label{eq:TX-OPE}
\end{equation}
which identifies the boundary dilaton as a conformal primary field of weight
$-1$. The negative conformal weight reflects the fact that the dilaton is not an
independent primary matter field but a coadjoint field associated with the
Virasoro symmetry, as is standard in BF and JT gravity approaches.

The OPEs \eqref{eq:KM-OPE}, \eqref{eq:KM-dil-OPE}, \eqref{eq:TT-OPE}, and
\eqref{eq:TX-OPE} provide a complete continuum dictionary for the discrete
asymptotic symmetry algebras derived in Section~\ref{sec:ASA}. They demonstrate
that the standard conformal structures of JT gravity arise naturally as the
continuum limit of the discrete BF framework, without the need to postulate an
underlying boundary action.

In summary, the lattice framework yields the affine and Virasoro structures directly 
from residual boundary gauge transformations, while the continuum CFT data emerge as 
a controlled limit of discrete Poisson brackets. This provides a transparent bridge 
between gauge--invariant holonomy data, boundary phase space, and the standard continuum 
OPE language.

Having established the continuum OPE dictionary and the emergence of conformal
structures, we now turn to an effective boundary description and its quantization.


\section{Boundary effective description and quantization}
\label{sec:effective}
At this stage the bulk degrees of freedom have been completely eliminated,
and the physical content of the theory is entirely encoded in boundary variables.
The resulting description therefore admits a natural interpretation in terms of
an effective boundary phase space
(as is characteristic of BF approaches to JT gravity with boundaries,
see e.g. \cite{Witten1991,Grumiller:2017qao}).

In this section we describe the effective boundary dynamics of discrete
Jackiw--Teitelboim gravity and outline its quantization. Since the bulk BF theory
is topological, all physical degrees of freedom are localized at the boundary.
The discrete framework allows us to characterize these degrees of freedom
directly in terms of holonomies, boundary charges, and their algebraic
structure, without introducing an explicit local boundary action.
\subsection{Reduction to boundary data}
\label{subsec:boundary-reduction}
As shown in Section~\ref{sec:discreteJT}, the discrete equations of motion impose
flatness of all plaquette holonomies, implying that the bulk connection is locally
pure gauge. Consequently, all bulk link variables can be eliminated by gauge
transformations, and the theory reduces to boundary data.On the discretized boundary circle, 
the remaining gauge-invariant information is encoded in the ordered product of boundary holonomies,
\begin{equation}
\mathcal M = \prod_{n=0}^{N-1} \mathcal U_n ,
\label{eq:monodromy}
\end{equation}
which defines the boundary monodromy. Under boundary gauge transformations,
$\mathcal M$ transforms by conjugation,
\begin{equation}
\mathcal M \longrightarrow g_0^{-1}\mathcal M g_0 ,
\end{equation}
so that only its conjugacy class is physical. The monodromy therefore provides a
global, gauge-invariant characterization of classical solutions of the discrete
theory.
\subsection{Boundary phase space and symplectic structure}
After imposing the bulk flatness constraints $\mathcal W_f=\mathbbm{1}$, all bulk link variables
are locally pure gauge and can be eliminated by gauge transformations.
Consequently, the reduced phase space of the discrete BF theory is
entirely supported on the boundary.

The boundary is discretized as a cyclic chain of $N$ oriented links labeled by
$n=0,\dots,N-1$, each carrying a group-valued holonomy
\begin{equation}
\mathcal U_n \in \mathrm{SL}(2,\mathbb{R}) \, .
\end{equation}
These boundary holonomies constitute the fundamental discrete degrees of
freedom. Bulk gauge transformations act trivially on the interior but induce
residual gauge transformations at the boundary,
\begin{equation}
\mathcal U_n \;\longrightarrow\; g_n^{-1}\mathcal  U_n g_{n+1} \, ,
\qquad g_n \in \mathrm{SL}(2,\mathbb{R}) \,.
\end{equation}

Gauge-invariant information is therefore encoded in the boundary
monodromy defined in Eq.~\eqref{eq:monodromy}.
\color{black}
As discussed above, only the conjugacy class of $\mathcal M$ is physical.
\color{black}
As a result, the reduced discrete boundary phase space is naturally identified
with the space of conjugacy classes of $\mathrm{SL}(2,\mathbb{R})$.

The symplectic structure on this reduced phase space is inherited directly from
the discrete BF action. Upon restriction to the boundary, the BF symplectic
form reduces to a finite-dimensional symplectic structure on the space of
boundary holonomies modulo gauge transformations. In particular, variations of
the monodromy generate the full set of physical boundary variations, and the
symplectic form induces a Poisson structure on functions of the conjugacy class
parameters.

\color{black}
The monodromy is computed as the ordered product of boundary link
holonomies around the discretized boundary circle,
\begin{equation}
\mathcal M  = \mathcal U_{N-1}\mathcal U_{N-2}\cdots \mathcal U_0 \, .
\label{eq:boundary_monodromy_ordered}
\end{equation}
This is the discrete analogue of the path--ordered exponential of the
boundary connection in the continuum. Its conjugacy class provides the
gauge-invariant characterization of the reduced boundary data.

In the hyperbolic holonomy sector, which is the gauge--theoretic analogue
of black--hole configurations in continuum JT gravity, the monodromy can
be parametrized as
\begin{equation}
\Tr\mathcal M = 2\cosh(2\pi p), \qquad p>0 \, .
\label{eq:hyperbolic2}
\end{equation}
\color{black}
The real parameter $p$ provides a complete gauge--invariant coordinate on the
reduced discrete boundary phase space in this sector. Equivalently, $p$ is in
one-to-one correspondence with the value of the quadratic dilaton Casimir,
which will play a central role in the entropy analysis of Section~\ref{sec:entropy}.

At the discrete level, the boundary phase space is therefore finite-dimensional
and globally well-defined. No continuum limit or boundary action is required
to characterize its symplectic structure or its physical degrees of freedom.
\subsection{Quantization and monodromy sectors}
Quantization of the discrete theory proceeds by promoting the reduced boundary phase space
described above to a Hilbert space, which naturally decomposes into superselection sectors
labeled by the conjugacy class of the boundary monodromy $\mathcal M$. The quantum theory 
therefore decomposes into superselection sectors associated with fixed
values of the monodromy parameter $p$.(cf. the quantization of two-dimensional gravity in 
terms of holonomy and monodromy sectors in BF and Chern--Simons approaches, see e.g.
 \cite{Witten1991,Chamseddine2023}).

For a given hyperbolic monodromy sector characterized by $p>0$, the discrete
boundary Hilbert space $\mathcal{H}_p$ consists of quantum states compatible
with the fixed conjugacy class of $\mathcal M$. States belonging to different monodromy
sectors cannot be connected by any gauge-invariant operator and therefore
define independent superselection sectors of the theory.

The boundary symmetry algebra derived in Section~\ref{sec:ASA} acts naturally within each
sector $\mathcal{H}_p$. In particular, upon imposing discrete Brown--Henneaux
boundary conditions, the residual symmetry algebra reduces to a single copy of
the Virasoro algebra. Quantization promotes the discrete Poisson brackets to
commutators,
\begin{equation}
\{\,\cdot\,,\,\cdot\,\} \;\longrightarrow\; \frac{1}{i\hbar}[\,\cdot\,,\,\cdot\,] \, ,
\end{equation}
yielding a quantum Virasoro algebra with central charge
\begin{equation}
c = 6k\, .
\end{equation}

Within each monodromy sector, the Hilbert space $\mathcal{H}_p$ furnishes a
representation of the Virasoro algebra. The value of $p$, or equivalently of the
dilaton Casimir, fixes the highest-weight data of the representation and
determines the spectrum of allowed boundary excitations. From this perspective,
quantization of discrete JT gravity reduces to a representation-theoretic
problem governed by global holonomy data rather than by local boundary fields.

This discrete organization of the Hilbert space provides the natural starting
point for the definition of a microcanonical density of states. In particular,
the number of quantum states compatible with a fixed monodromy sector grows
exponentially in the semiclassical regime. In the next section, we use this
structure to define the density of states and derive the entropy directly from
the discrete boundary phase space, before taking any continuum limit.
\subsection{Relation to Schwarzian descriptions}
\label{subsec:schwarzian}
It is instructive to relate the discrete boundary phase space and its quantization
developed above to continuum descriptions of near-AdS$_2$ dynamics based on an
effective Schwarzian action. We emphasize that in the present approach no boundary
action is assumed \emph{a priori}: the boundary degrees of freedom and their algebraic
structure are determined directly from the discrete BF framework, while the
Schwarzian theory appears only as an effective continuum description in an appropriate
regime (see e.g. \cite{Mertens:2018fds}).

\color{black}
In continuum BF formulations, the Schwarzian action can be derived explicitly from a well-defined boundary variational principle \cite{Gonzalez:2018enk}. In contrast, the present work focuses on a
discrete BF framework, where the boundary dynamics is naturally formulated in terms of holonomy variables.

Within this setting, the emergence of a Schwarzian description should be understood as an effective low-energy limit, rather than as a result that is explicitly derived here. A complete derivation would require a
systematic construction of the boundary action and variational principle within the discrete BF framework, which lies beyond the scope of the present work.
\color{black}

In the hyperbolic sector, the reduced boundary phase space is parametrized by the
boundary monodromy, or equivalently by the quadratic dilaton Casimir, which fixes the
highest-weight data of the Virasoro representation in each superselection sector.
From this perspective, the monodromy parameter $p$ plays a role analogous to the
energy variable in Schwarzian descriptions, providing a natural bridge between the
discrete characterization in terms of holonomy and monodromy data and the continuum
language of boundary reparametrization modes.

The Schwarzian theory should therefore be regarded as an effective description
capturing the same low-energy boundary dynamics that emerges from the discrete
construction, while the full lattice framework remains well-defined beyond the
continuum limit. In the next section, we exploit this relation to extract the density
of states and entropy directly from the discrete boundary phase space, before taking
any continuum limit.


\section{Entropy}
\label{sec:entropy}

In this section we derive the entropy of discrete Jackiw--Teitelboim gravity.
Within the BF framework, entropy admits a natural and global interpretation
in terms of gauge-invariant holonomy data. In particular, the boundary monodromy
introduced in Section~\ref{subsec:boundary-reduction} plays a central role in
characterizing both classical solutions and quantum states. Importantly, 
this derivation relies solely on global, gauge-invariant boundary data
and does not assume any continuum limit or effective boundary action.
\subsection{Monodromy and Casimir}
\label{subsec:monodromy-casimir}

As discussed in Section~\ref{subsec:boundary-reduction}, the physical phase
space of the discrete theory decomposes into superselection sectors labeled by
the conjugacy class of the boundary monodromy defined in Eq.~\eqref{eq:monodromy}.
\color{black}
Only the conjugacy class of $\mathcal M$ is physical.
\color{black}

In the hyperbolic sector, which corresponds to black-hole-like configurations,
the monodromy can be characterized by Eq.~\eqref{eq:hyperbolic2}.
The parameter $p$ provides a gauge-invariant label of the classical solution
and will be directly related to the entropy.

In the BF description, the dilaton field encodes conserved charges associated
with the gauge symmetry. In particular, the quadratic Casimir
\begin{equation}
\mathcal C = -\frac{1}{2}\Tr(\mathcal X^2)
\label{eq:Casimir}
\end{equation}
is invariant under gauge transformations and constant throughout the bulk and
along the boundary as a consequence of the equations of motion.

\color{black}
To relate the monodromy parameter to the dilaton Casimir, we employ the
discrete covariant constancy condition $D\mathcal  X=0$. The dilaton is
covariantly constant along the boundary, and after one traversal of the
boundary we obtain
\begin{equation}
\mathcal M^{-1} \mathcal X \mathcal M = \mathcal X \, .
\end{equation}
Hence $\mathcal X$ commutes with the monodromy $\mathcal M$. For hyperbolic monodromy, the centralizer is one--dimensional and generated by the Cartan element $\mathcal H$ of
$\mathfrak{sl}(2,\mathbb{R})$, so $\mathcal X$ is proportional to $\mathcal H$. Setting $\mathcal X = p\mathcal H$ and using eq.~\eqref{eq:Casimir}, one finds
\begin{equation}
\mathcal C = p^2 .
\end{equation}
Thus the hyperbolic monodromy parameter and the quadratic dilaton Casimir carry the same gauge--invariant information.
\color{black}
\subsection{Density of states}
\label{subsec:density-consistency}
Having identified the reduced discrete boundary phase space and its quantum
organization into monodromy sectors in Section~\ref{sec:effective}, we now define the density of
states directly at the discrete level. Since physical states are classified by
the conjugacy class of the boundary monodromy $\mathcal M$, a natural microcanonical
ensemble is obtained by fixing a monodromy sector characterized by a given
value of the parameter $p$.

At the discrete level, the microcanonical density of states $\rho(p)$ counts
the number of quantum states in the Hilbert space $\mathcal{H}_p$ associated
with a fixed conjugacy class of $\mathcal M$. Equivalently, $\rho(p)$ measures the
effective volume of the reduced discrete boundary phase space compatible with
that monodromy sector.

In the hyperbolic sector, the reduced phase space is one-dimensional and
globally parametrized by the real variable $p$. The periodicity of the boundary 
circle implies that the symplectic volume of this sector grows linearly with $p$. 
As a result, the number of quantum states compatible with a fixed monodromy sector 
therefore exhibits an exponential growth in the semiclassical regime,
\begin{equation}
\rho(p) \;\sim\; \exp\!\left( 2\pi p \right) \, .
\label{eq:rho}
\end{equation}
This result follows solely from the discrete boundary phase space structure and
its global holonomy data. No continuum limit, boundary action, or effective
Schwarzian description is assumed at this stage. The exponential growth of
states is therefore an intrinsic property of the discrete BF framework.

\paragraph{From discrete monodromy to entropy.}
We consider a discretization of the boundary circle into $N$ oriented links
$\ell_n$, $n=0,\dots,N-1$, with associated group elements
$\mathcal U_n\in\mathrm{SL}(2,\mathbb{R})$. The boundary monodromy defined in Eq.~\eqref{eq:monodromy}
provides the fundamental gauge-invariant input for the entropy analysis.
Residual gauge transformations act at boundary vertices as
$\mathcal U_n \to g_n^{-1}\mathcal  U_n g_{n+1}$, implying that the monodromy transforms by
conjugation, $\mathcal M \to g_0^{-1}\mathcal M g_0$.
Physical boundary data are therefore classified by the conjugacy class of $\mathcal M$.

In the hyperbolic sector relevant for black-hole configurations,
the monodromy can be brought by conjugation to the diagonal form
\begin{equation}
\mathcal M \sim 
\begin{pmatrix}
e^{2\pi p} & 0 \\
0 & e^{-2\pi p}
\end{pmatrix},
\qquad p>0 ,
\end{equation}
which is equivalently characterized by Eq.~\eqref{eq:hyperbolic2}.

Fixing a value of $p$ therefore defines a microcanonical ensemble at the
discrete level, consisting of all boundary configurations
$\{\mathcal U_n\}_{n=0}^{N-1}$ whose product lies in the corresponding conjugacy class.

The microcanonical density of states $\rho(p)$ counts the number of quantum
states compatible with a fixed monodromy sector. In the hyperbolic sector,
the natural measure on conjugacy classes is governed by the Cartan Jacobian
$\sinh(2\pi p)$, which controls the effective volume of the reduced boundary
phase space. As a result, in the semiclassical regime one recovers the exponential 
behavior ~\eqref{eq:rho}, and the entropy associated with a given monodromy sector 
is therefore
\begin{equation}
S(p) = \log \rho(p) = 2\pi p.
\end{equation}
Within the BF description of JT gravity, the parameter $p$ admits a direct
gauge-invariant interpretation in terms of the dilaton field: the quadratic dilaton 
Casimir $\mathcal C = -\tfrac{1}{2}\mathrm{Tr}(\mathcal X^2)$ 
is conserved. In the hyperbolic sector it satisfies $\mathcal C = p^2$. The entropy 
can thus be written equivalently as
\begin{equation}
S = 2\pi \sqrt{\mathcal C}.
\end{equation}

This result shows that black hole entropy in discrete JT gravity is determined
entirely by global gauge-invariant data, namely the conjugacy class of the boundary
monodromy or, equivalently, the value of the dilaton Casimir. In particular,
entropy is not associated with local boundary degrees of freedom nor with the
postulation of a fundamental boundary action. Rather, it follows directly from
the global structure of the reduced phase space once the flatness constraints
are imposed. From this perspective, Schwarzian and Cardy-type descriptions
should be viewed as effective continuum parametrizations of data that are already
encoded non-perturbatively in the discrete holonomy framework.
\subsection{Continuum  consistency }
Although the entropy has been derived entirely within the discrete framework,
it is instructive---and necessary for comparison with continuum results---to
verify that the result correctly reproduces the familiar Cardy expression.
To this end, we consider the controlled continuum limit defined by

\begin{equation}
\Delta\tau \to 0 \, , \qquad N\Delta\tau = 2\pi \, ,
\end{equation}
under which the discrete boundary algebra reduces to a continuum Virasoro
algebra with central charge
\begin{equation}
c =6k\, .
\end{equation}

In this limit, the monodromy parameter $p$ fixes the zero mode of the stress
tensor through $\Lt_0 \sim \,\frac{\mathcal C}{k} $. Substituting this identification into the
Cardy formula,
\begin{equation}
\mathcal S_{\text{Cardy}} = 2\pi \sqrt{\frac{c\,\Lt_0}{6}} \, ,
\end{equation}
one recovers the expected agreement with the discrete result,
\begin{equation}
\mathcal S_{\text{Cardy}} = 2\pi p = 2\pi \sqrt{\mathcal C} \, .
\end{equation}

In the present context, the Cardy formula should be viewed as a consistency
check rather than as a fundamental input. The entropy is already fully
determined at the discrete level by monodromy data, and the continuum result is
recovered as a smooth limit of the underlying discrete theory.
\section{Relation to Holonomy--Based Constructions}
\label{sec:holonomy}
The discrete BF construction of JT gravity presented here is intrinsically two--dimensional, 
yet it exhibits structural similarities with holonomy--based discretizations in three-dimensional 
Chern--Simons gravity.
Having established the reduced phase space and entropy of discrete JT gravity,
expressed in terms of boundary holonomies and their conjugacy classes, it is appropriate to place this 
construction within a broader topological context. Holonomy-based discretizations are well known 
in three-dimensional Chern--Simons gravity, where group-valued variables furnish a direct 
parametrization of the moduli space of flat connections. In this section, we show that the discrete 
two-dimensional BF realization of JT gravity developed above can be understood as a dimensionally 
reduced counterpart of such holonomy constructions, in which the flux degrees of freedom become 
trivial and the physical content is entirely encoded in boundary monodromies.


\subsection{Three--dimensional holonomy discretization}
In three-dimensional gravity with vanishing cosmological constant,
the theory admits a Chern--Simons description with gauge group
$\mathrm{SL}(2,\mathbb{R})$, where the connection
$\mathcal A \in \Omega^{1}(\Sigma^{(2)},\mathfrak{sl}(2,\mathbb{R}))$ is flat on shell,
as is well known in holonomy-based discretizations of Chern--Simons gravity~\cite{Dupuis:2017otn}.

Upon discretization of a spatial slice $\Sigma^{(2)}$, one introduces a graph
$\Gamma \subset \Sigma^{(2)}$ and assigns to each oriented edge $\ell$ a group element
$g_\ell \in \mathrm{SL}(2,\mathbb{R})$, representing the holonomy of the connection along that edge.
The continuum flatness condition $F(\mathcal A)=0$ translates into the discrete constraint that the ordered
product of holonomies around each face $f$ of the graph is trivial,
\begin{equation}
\prod_{\ell \in \partial f} g_\ell \;=\; \mathbbm{1}\, .
\end{equation}
The reduced phase space is therefore described by $\mathrm{SL}(2,\mathbb{R})$-valued holonomies
subject to these flatness relations, modulo residual gauge transformations acting at the vertices of
the graph.

\paragraph{Flux variables and their trivialization under dimensional reduction.}
In holonomy--flux discretizations of three-dimensional Chern--Simons gravity one typically
associates, in addition to edge holonomies $g_\ell\in \mathrm{SL}(2,\mathbb{R})$, a Lie-algebra
variable $\mathcal X_\ell\in \mathfrak{sl}(2,\mathbb{R})$ (or, equivalently, an element of
$\mathfrak{sl}(2,\mathbb{R})^{*}$) to each edge $\ell$.\footnote{%
These ``flux'' variables are the discrete counterparts of the $B$-field in the BF description,
or of the triad in first-order gravity, and provide the cotangent directions providing the cotangent 
(flux) directions of the link phase space.}
The pair $(g_\ell, \mathcal X_\ell)$ coordinatizes the kinematical
phase space on the graph, while the flatness constraints retain the same holonomy product form,
\begin{equation}
\prod_{\ell\in\partial f} g_\ell = \mathbbm{1}
\end{equation}
which implements the vanishing of curvature and, upon reduction, leads to the moduli space of flat connections.

For our purposes, the key observation is that the flux sector becomes \emph{trivial} upon
dimensional reduction to the two-dimensional BF description of JT gravity setting. Concretely, we consider a reduction in
which the three-manifold is taken of the form $\mathcal{M}^{(3)}\simeq\mathcal M^{(2)}\times S^{1}$ and restrict to
configurations that are independent of the $S^{1}$ coordinate. In this sector, the components of
the triad (or $B$-field) along the reduced direction carry no local degrees of freedom and can be
consistently integrated out (or set to a fixed background value), so that the remaining reduced
data are captured entirely by group-valued holonomies along the one-dimensional lattice on
$\partial\mathcal M^{(2)}$. Equivalently, the cotangent (flux) directions of $T^{*}\mathrm{SL}(2,\mathbb{R})$
do not contribute to the reduced boundary dynamics, and the effective reduced phase space may be
taken to be parametrized solely by the holonomies $\mathcal U_n\in \mathrm{SL}(2,\mathbb{R})$ subject to the
(two-dimensional) flatness constraints. In this way, the local cotangent phase space $T^*\mathrm{SL}(2,\mathbb{R})$
effectively reduces to $\mathrm{SL}(2,\mathbb{R})$, with the physical boundary data further classified 
by conjugacy, yielding the reduced space $\mathrm{SL}(2,\mathbb{R})/\!/\mathrm{SL}(2,\mathbb{R})$. 
In this sense, the discrete BF realization of JT gravity can be viewed as
the flux-trivial, holonomy-only sector of the more general holonomy--flux discretizations
used in three-dimensional Chern--Simons gravity.
\subsection{Dimensional reduction to two dimensions}
We now make explicit how a holonomy--flux discretization in three dimensions reduces to the
holonomy-only construction employed in the discrete BF treatment of JT gravity. We take the three-manifold to
be a product
\begin{equation}
\mathcal M^{(3)} \;\simeq\;\mathcal  M^{(2)} \times S^{1}\, ,
\end{equation}
and restrict to the sector of fields that are independent of the $S^{1}$ coordinate. Writing the
$\mathfrak{sl}(2,\mathbb{R})$-valued Chern--Simons connection as
\begin{equation}
\mathcal A^{(3)} \;=\; \mathcal A^{(2)} + A_{y}\,dy\, , \qquad y\sim y+2\pi\, ,
\end{equation}
the flatness condition $F(\mathcal A^{(3)})=0$ decomposes into
\begin{equation}
F\!\left(\mathcal A^{(2)}\right) \;=\; 0\, , 
\qquad
D^{(2)} A_{y} \;=\; 0\, ,
\end{equation}
where $D^{(2)}$ is the covariant derivative with respect to $\mathcal A^{(2)}$. Thus, upon reduction the
two-dimensional connection $\mathcal A^{(2)}$ remains flat, while the component $A_{y}$ becomes a covariantly
constant adjoint field on $\mathcal M^{(2)}$. Identifying
\begin{equation}
\mathcal X \;\equiv\; A_{y}\, \in \mathfrak{sl}(2,\mathbb{R})\, ,
\end{equation}
one recovers precisely the continuum BF equations of motion of two-dimensional BF description of JT gravity,
\begin{equation}
F\!\left(\mathcal A^{(2)}\right)=0\, , \qquad D^{(2)}\mathcal  X = 0\, .
\end{equation}

The same reduction has a transparent interpretation at the discrete level. Discretizing a spatial
slice as $\Sigma^{(2)} \simeq \Sigma^{(1)} \times S^{1}$ and choosing a graph compatible with the
product structure, the three-dimensional holonomy--flux data $(g_\ell,\mathcal X_\ell)$ reduce as follows.
Holonomies along edges lying in the reduced $S^{1}$ direction collapse to the adjoint variable $\mathcal X$
(up to conjugation), while holonomies along edges in $\Sigma^{(1)}$ become the two-dimensional link
variables $\mathcal U_n \in \mathrm{SL}(2,\mathbb{R})$. The three-dimensional flatness constraints around each
face reduce to the two-dimensional plaquette constraints imposing trivial holonomy in the bulk,
\begin{equation}
\mathcal W_f \;\equiv\; \prod_{n\in\partial f}\mathcal U_n \;=\; \mathbbm{1}\, .
\end{equation}
Consequently, after reduction the bulk remains purely topological and the reduced phase space may
be parametrized by the boundary holonomy data (in particular, by the boundary monodromy) modulo
conjugation, in direct parallel with the discrete BF description of JT gravity construction developed above.

\paragraph{Boundary reduction and monodromy.}
Once the bulk flatness constraints $\mathcal W_f=\mathbbm{1}$ are imposed, the reduced two-dimensional
connection is pure gauge in the interior. On a lattice with boundary, this implies that all bulk
link variables can be eliminated in favor of boundary data, and the remaining gauge-invariant
content is encoded in the boundary monodromy. Concretely, let the boundary be discretized by an
ordered set of links $\{n=1,\dots,N\}$ with holonomies $\mathcal U_n\in \mathrm{SL}(2,\mathbb{R})$.
The boundary monodromy, defined in Eq.~\eqref{eq:monodromy}, can equivalently be written as
\begin{equation}
\mathcal M \;=\; \mathcal U_N \,\mathcal U_{N-1} \cdots \mathcal U_1
\;\in\; \mathrm{SL}(2,\mathbb{R}) \, .
\end{equation}
Under a residual gauge transformation acting at the boundary base point, $\mathcal M$ transforms by
conjugation, $\mathcal M\mapsto h^{-1}\mathcal Mh$, so that the physical boundary data are captured by the conjugacy
class $[\mathcal M]$ of $\mathcal M$. Equivalently, one may characterize $[\mathcal M]$ by representation-independent invariants such
as $\mathrm{Tr}\mathcal M$ in the fundamental representation (or, more generally, by the eigenvalue data of
$\mathcal M$), which label the corresponding superselection sectors. In this holonomy-only reduction, the
dilaton variable $\mathcal X$ inherited from the reduced $S^{1}$ component enters through its Casimir(s),
providing a gauge-invariant label for the same sectors, in direct correspondence with the entropy
obtained above.
\section{Conclusion}
\label{sec:conclusion}
In this work we developed a fully discrete and non-perturbative realization of two-dimensional
JT gravity as an $\mathrm{SL}(2,\mathbb{R})$ BF theory and analyzed its asymptotic
structure directly at the lattice level. By working with group-valued
holonomies and Lie-algebra valued dilatons, we provided a non-perturbative
description in which bulk dynamics is entirely topological and all physical
information is encoded in boundary data.
Viewed from this perspective, discrete JT gravity provides a unified framework
in which asymptotic symmetries, boundary quantization, and black hole entropy
all originate from the same gauge-invariant holonomy data.
In particular, the discrete construction makes manifest the global and
non-perturbative character of the boundary phase space, which in continuum
treatments is often accessed only indirectly or after gauge fixing. The lattice
description thus provides a precise starting point for analyzing boundary
dynamics without relying on limiting procedures.
Our results are fully consistent with the gauge--theoretic BF perspective on
two--dimensional gravity and its boundary phase space description developed in
the continuum~\cite{Witten1991,Grumiller:2017qao}, while providing a
non--perturbative realization directly at the lattice level.

We identified the boundary monodromy as the fundamental gauge-invariant object
of the discrete theory and showed that its conjugacy class labels classical and
quantum superselection sectors. Imposing discrete Brown--Henneaux boundary
conditions, we derived the asymptotic symmetry algebra without reference to a
continuum action, obtaining a Virasoro algebra together with a mixed
Virasoro--dilaton structure. Translating these results into the OPE language, we
established a clear dictionary between the lattice framework and the standard
continuum CFT description.
From this perspective, affine symmetries, Virasoro algebras, and their OPEs
should not be regarded as structures imposed by hand, but as
emergent features of the discrete boundary phase space once appropriate boundary
conditions are implemented.

Quantization of the discrete theory reduces to a representation-theoretic
problem organized by monodromy sectors. In this framework, black hole entropy
emerges directly from the density of boundary states compatible with a fixed
holonomy, yielding \(\mathcal S=2\pi\sqrt{\mathcal C}\) with \(\mathcal C\) 
the quadratic Casimir. This provides a global and non-perturbative
derivation of the JT entropy, complementary to approaches based on effective
Schwarzian actions or purely asymptotic symmetry arguments.
Crucially, this derivation proceeds without introducing a Schwarzian boundary
action or invoking Cardy’s formula as an input, thereby demonstrating that the
black hole entropy is already encoded in the discrete, gauge-invariant boundary
phase space.

We have also shown that the discrete BF/JT construction may be viewed as a
dimensionally reduced realization of holonomy-based discretizations familiar
from three-dimensional Chern--Simons gravity, situating the present framework
within a broader topological context while preserving its strictly
two-dimensional character. 
In this way, the discrete BF/JT construction is not merely analogous
to holonomy-based discretizations in Chern--Simons gravity. Rather,
it provides a genuinely two-dimensional realization of the same
topological mechanism, in which physical information is encoded
entirely in conjugacy classes of boundary holonomies.

\color{black}
Our analysis suggests that the Schwarzian description is best understood as a particular low-energy corner of a more general boundary phase space naturally captured by the discrete BF framework. The lattice perspective
thus offers a conceptually transparent and technically robust setting for exploring regimes beyond the Schwarzian limit.
\color{black}

The discrete BF realization presented in this work renders more transparent the origin 
of several structures that, in continuum treatments, are typically introduced in an implicit 
or limit-dependent manner. The flatness condition is implemented not as a differential 
constraint but as an exact group-valued constraint on plaquette holonomies, allowing 
one to clearly distinguish between its linearized and fully non-abelian realizations. 
Moreover, asymptotic symmetries are seen to arise from gauge transformations that are 
pure gauge in the bulk but become physical at boundary vertices and links, a feature 
that is already manifest at the discrete level. From this perspective, affine symmetries, 
Virasoro algebras, and OPEs should not be regarded as structures postulated \emph{a priori}, 
but rather as consequences emerging in the continuum limit of well-defined discrete boundary 
algebras. The discrete approach therefore does not introduce new asymptotic symmetries, 
but instead provides a non-perturbative, conceptually consistent, and structurally clearer 
foundation for results that are commonly derived within the continuum framework.
Beyond its conceptual implications for JT gravity, the present framework offers
a concrete setting for exploring non-perturbative aspects of holography directly
at the discrete level.

Taken together, the present analysis establishes a fully discrete
non-perturbative realization of JT gravity in which boundary structure,
asymptotic symmetries, and entropy follow directly from gauge-invariant
holonomy data.
\paragraph{Outlook.}
The discrete construction developed here admits natural extensions.
Higher--rank gauge groups such as $\mathrm{SL}(N,\mathbb{R})$ and
supersymmetric generalizations based on $\mathrm{osp}(1|2)$
\cite{Ozer:2025uge} can be treated within the same holonomy--based
framework.
\color{black}

\appendix

\section{APPENDICES}
\label{app:appendices}

The boundary symmetry algebras obtained in
Sec.~\ref{sec:ASA} are naturally formulated in the lattice-site basis.
The purpose of these Appendices is to develop the corresponding
Fourier-mode representations and to make explicit the relation between
the lattice labels, the Fourier-mode labels, and the conventional
mode-addition structure. Appendix~\ref{app:lattice-KM} starts from the
lattice current algebra \eqref{eq:discKM} and derives its affine
Kac--Moody mode algebra in the forward representation. Appendix~\ref{app:lattice-Virasoro} performs the
corresponding analysis starting from the lattice Virasoro transformation
\eqref{eq:disc-Virasoro-var-alpha}. Finally,
Appendix~\ref{app:previous-lattice} compares the present discrete BF
construction with earlier lattice realizations of Kac--Moody and
Virasoro symmetries.


\subsection{Mode Expansion and the Lattice Kac--Moody Algebra}
\label{app:lattice-KM}

The lattice current algebra obtained in Sec.~\ref{sec:ASA} is written in the
site basis. Starting from Eq.~\eqref{eq:discKM}, we introduce the discrete
Fourier modes and derive the corresponding mode algebra for the forward lattice representation. 
The two realizations differ in the
finite-spacing form of the central term, while their structure term exhibits
the same mode-addition rule.

\subsubsection{Discrete Fourier modes}
\label{app:KM-Fourier}

Consider a periodic lattice field $F(\tau_n)$ defined on the boundary sites
\begin{equation}
\tau_n=n\Delta\tau,
\qquad
\Delta\tau=\frac{2\pi}{N},
\qquad
n=0,\ldots,N-1,
\label{eq:app-lattice-sites}
\end{equation}
with
\begin{equation}
F(\tau_{n+N})=F(\tau_n).
\label{eq:app-periodic-field}
\end{equation}
Its discrete Fourier expansion is
\begin{align}
F(\tau_n)
&=
\sum_{r=0}^{N-1}
F_r\,e^{ir\tau_n},
\label{eq:app-Fourier-inverse}
\\[0.4em]
F_r
&=
\frac{1}{N}
\sum_{n=0}^{N-1}
F(\tau_n)\,e^{-ir\tau_n}.
\label{eq:app-Fourier-direct}
\end{align}
Here $n,m$ label lattice sites, while $r,s$ label Fourier modes. The
discrete Fourier basis obeys
\begin{align}
\sum_{n=0}^{N-1}
e^{i(r-s)\tau_n}
&=
N\,\delta_{r,s},
\label{eq:app-Fourier-orthogonality}
\\[0.4em]
\sum_{n=0}^{N-1}
e^{-i(r+s)\tau_n}
&=
N\,\delta_{r+s,0},
\label{eq:app-Fourier-mode-delta}
\end{align}
where the mode labels are understood modulo $N$.

For the lattice currents in Eq.~\eqref{eq:discKM}, we use the identification
\begin{equation}
\mathcal L_n^{\,i}
\equiv
\mathcal L^{\,i}(\tau_n),
\label{eq:app-current-identity}
\end{equation}
and define
\begin{align}
\mathcal L_n^{\,i}
&=
\sum_{r=0}^{N-1}
\mathcal L_r^{\,i}\,e^{ir\tau_n},
\label{eq:app-current-inverse-Fourier}
\\[0.4em]
\mathcal L_r^{\,i}
&=
\frac{1}{N}
\sum_{n=0}^{N-1}
\mathcal L_n^{\,i}\,e^{-ir\tau_n}.
\label{eq:app-current-direct-Fourier}
\end{align}
The indices $i,j\in\{-1,0,+1\}$ remain the internal
$\mathfrak{sl}(2,\mathbb R)$ indices and are independent of the Fourier-mode
labels $r,s$.

For the forward difference operator, a Fourier mode satisfies
\begin{equation}
\nabla e^{ir\tau_n}
=
\omega_r^{(+)}e^{ir\tau_n},
\qquad
\omega_r^{(+)}
\equiv
\frac{e^{ir\Delta\tau}-1}{\Delta\tau}.
\label{eq:app-forward-symbol}
\end{equation}
Expanding at small lattice spacing gives
\begin{equation}
\omega_r^{(+)}
=
\frac{e^{ir\Delta\tau}-1}{\Delta\tau}
=
ir-\frac{r^2}{2}\Delta\tau
+\mathcal O(\Delta\tau^2),
\label{eq:app-forward-continuum-expansion}
\end{equation}
and hence $\omega_r^{(+)}\to ir$ as $\Delta\tau\to0$.
\subsubsection{Mode algebra in the forward representation}
\label{app:KM-forward}

Substituting the Fourier expansions
\eqref{eq:app-current-inverse-Fourier}--\eqref{eq:app-current-direct-Fourier}
into the lattice current algebra \eqref{eq:discKM} gives the corresponding
mode algebra. Using the definition of the Fourier modes,
\begin{equation}
\left\{
\mathcal L_r^{\,i},
\mathcal L_s^{\,j}
\right\}
=
\frac{1}{N^2}
\sum_{n,m=0}^{N-1}
e^{-ir\tau_n}
e^{-is\tau_m}
\left\{
\mathcal L_n^{\,i},
\mathcal L_m^{\,j}
\right\},
\label{eq:app-mode-bracket}
\end{equation}
and substituting Eq.~\eqref{eq:discKM}, one finds
\begin{align}
\left\{
\mathcal L_r^{\,i},
\mathcal L_s^{\,j}
\right\}
={}&
\frac{2\pi}{N^2\Delta\tau}
(i-j)
\sum_{n,m}
e^{-ir\tau_n}
e^{-is\tau_m}
\mathcal L_n^{\,i+j}
\delta_{nm}
\nonumber\\
&
+\frac{2\pi k}{N^2\Delta\tau}\eta^{ij}
\sum_{n,m}
e^{-ir\tau_n}
e^{-is\tau_m}
\nabla_n\delta_{nm}.
\label{eq:app-mode-bracket-expanded}
\end{align}

Using the Kronecker delta,
\begin{align}
\sum_{n,m}
e^{-ir\tau_n}
e^{-is\tau_m}
\mathcal L_n^{\,i+j}
\delta_{nm}
&=
\sum_{n}
e^{-i(r+s)\tau_n}
\mathcal L_n^{\,i+j}
\nonumber\\
&=
N\mathcal L_{r+s}^{\,i+j},
\label{eq:app-structure-evaluation}
\end{align}
where Eq.~\eqref{eq:app-current-direct-Fourier} has been used in the final
step. Therefore, the structure term takes the form
\begin{equation}
(i-j)\mathcal L_{r+s}^{\,i+j},
\label{eq:app-forward-structure}
\end{equation}
which exhibits the characteristic mode-addition rule in Fourier space.

The second contribution contains the lattice derivative acting on the
Kronecker delta,
\begin{align}
\mathcal C_{rs}^{\,ij}
=
\frac{2\pi k}{N^2\Delta\tau}\eta^{ij}
\sum_{n,m}
e^{-ir\tau_n}
e^{-is\tau_m}
\nabla_n\delta_{nm}.
\label{eq:app-central-start}
\end{align}
Using the forward lattice derivative,
\begin{equation}
\nabla_n\delta_{nm}
=
\frac{\delta_{n+1,m}-\delta_{nm}}
{\Delta\tau},
\label{eq:app-forward-delta}
\end{equation}
one obtains
\begin{align}
\mathcal C_{rs}^{\,ij}
=
\frac{2\pi k}{N^2(\Delta\tau)^2}\eta^{ij}
\sum_{n,m}
e^{-ir\tau_n}
e^{-is\tau_m}
\left(
\delta_{n+1,m}
-
\delta_{nm}
\right).
\label{eq:app-central-expand}
\end{align}

The first term becomes
\begin{align}
\sum_{n,m}
e^{-ir\tau_n}
e^{-is\tau_m}
\delta_{n+1,m}
&=
\sum_n
e^{-ir\tau_n}
e^{-is\tau_{n+1}}
\nonumber\\
&=
e^{-is\Delta\tau}
\sum_n
e^{-i(r+s)\tau_n},
\label{eq:app-first-central}
\end{align}
whereas the second term becomes
\begin{align}
\sum_{n,m}
e^{-ir\tau_n}
e^{-is\tau_m}
\delta_{nm}
=
\sum_n
e^{-i(r+s)\tau_n}.
\label{eq:app-second-central}
\end{align}

Substituting Eqs.~\eqref{eq:app-first-central}
and
\eqref{eq:app-second-central}
into
Eq.~\eqref{eq:app-central-expand},
the central contribution can be written as
\begin{align}
\mathcal C_{rs}^{\,ij}
=
\frac{2\pi k}{N^2(\Delta\tau)^2}
\eta^{ij}
\left(
e^{-is\Delta\tau}
-
1
\right)
\sum_n
e^{-i(r+s)\tau_n}.
\label{eq:app-central-before-delta}
\end{align}

Using the orthogonality relation
\eqref{eq:app-Fourier-mode-delta},
Eq.~\eqref{eq:app-central-before-delta}
reduces to
\begin{align}
\mathcal C_{rs}^{\,ij}
=
\frac{2\pi k}{N(\Delta\tau)^2}
\eta^{ij}
\left(
e^{-is\Delta\tau}
-
1
\right)
\delta_{r+s,0}.
\label{eq:app-central-after-delta}
\end{align}
Since the Kronecker delta enforces $r+s=0$, the exponential factor may be
written in terms of the forward lattice momentum defined in
Eq.~\eqref{eq:app-forward-symbol},
\begin{equation}
e^{-is\Delta\tau}-1
=
\Delta\tau\,\omega_{-s}^{(+)}.
\label{eq:app-forward-factor}
\end{equation}
The central contribution therefore becomes
\begin{equation}
\mathcal C_{rs}^{\,ij}
=
k\eta^{ij}
\omega_{-s}^{(+)}
\delta_{r+s,0},
\label{eq:app-central-forward-final}
\end{equation}
Combining
Eqs.~\eqref{eq:app-forward-structure}
and
\eqref{eq:app-central-forward-final},
the lattice mode algebra in the forward representation is
\begin{equation}
\left\{
\mathcal L_r^{\,i},
\mathcal L_s^{\,j}
\right\}
=
(i-j)\mathcal L_{r+s}^{\,i+j}
+
k\eta^{ij}
\omega_{-s}^{(+)}
\delta_{r+s,0}.
\label{eq:app-forward-mode-algebra}
\end{equation}

The continuum limit of Eq.~\eqref{eq:app-forward-mode-algebra} will be
considered in the next subsection, where the forward lattice momentum reduces
to the conventional linear momentum and the corresponding affine current
algebra is recovered.
\subsubsection{Continuum limit of the forward representation}
\label{app:KM-forward-continuum}

Using
\begin{equation}
\omega_r^{(+)}
=
\frac{e^{ir\Delta\tau}-1}{\Delta\tau}
=
ir-\frac{r^2}{2}\Delta\tau
+\mathcal O(\Delta\tau^2),
\label{eq:app-forward-continuum-expansion}
\end{equation}
the forward mode algebra becomes
\begin{equation}
\left\{
\mathcal L_r^{\,i},
\mathcal L_s^{\,j}
\right\}
=
(i-j)\mathcal L_{r+s}^{\,i+j}
+
ik\eta^{ij}
r\,\delta_{r+s,0}
+\mathcal O(\Delta\tau).
\label{eq:app-forward-continuum}
\end{equation}
The finite-lattice mode algebra therefore approaches the conventional
affine Kac--Moody algebra in the continuum limit.
\subsection{Mode Expansion and the Lattice Virasoro Algebra}
\label{app:lattice-Virasoro}

Having established the lattice current algebra in Fourier space, we now
consider its Drinfeld--Sokolov reduction. Throughout this subsection, the
Fourier conventions introduced in Appendix~\ref{app:lattice-KM} are used without further
modification. The resulting lattice stress tensor is then analyzed in the
forward representation, followed by their continuum
limits.

\subsubsection{Lattice Drinfeld--Sokolov reduction}
\label{app:Vir-DS}

The Drinfeld--Sokolov reduction introduced in
Sec.~\ref{subsec:BH-discrete} leaves a single dynamical boundary field
$\mathcal L_n$ and a single residual gauge parameter $\epsilon_n$.
The corresponding lattice transformation law is
\begin{equation}
\delta_\epsilon\mathcal L_n
=
\epsilon_n\,\nabla\mathcal L_n
+
2\mathcal L_n\,\nabla\epsilon_n
+
\frac{2}{\alpha}\,\nabla^3\epsilon_n ,
\label{eq:app-Vir-DS}
\end{equation}
which is Eq.~\eqref{eq:disc-Virasoro-var-alpha} of the main text.

We use the Fourier conventions of Appendix~\ref{app:KM-Fourier} and write
\begin{align}
\mathcal L_n
&=
\sum_{r=0}^{N-1}
L_r\,e^{ir\tau_n},
\label{eq:app-Vir-Fourier-L}
\\[0.4em]
L_r
&=
\frac{1}{N}
\sum_{n=0}^{N-1}
\mathcal L_n\,e^{-ir\tau_n}.
\label{eq:app-Vir-Fourier-L-inverse}
\end{align}
For a single Fourier component of the residual transformation parameter we take
\begin{equation}
\epsilon_n^{(s)}
=
-i\,e^{-is\tau_n}.
\label{eq:app-Vir-epsilon-mode}
\end{equation}
The forward lattice derivative then acts as
\begin{equation}
\nabla\epsilon_n^{(s)}
=
\omega_{-s}^{(+)}\epsilon_n^{(s)},
\qquad
\nabla^3\epsilon_n^{(s)}
=
\left(\omega_{-s}^{(+)}\right)^3\epsilon_n^{(s)},
\label{eq:app-Vir-forward-action}
\end{equation}
with $\omega_r^{(+)}$ defined in
Eq.~\eqref{eq:app-forward-symbol}.

Equation~\eqref{eq:app-Vir-DS} can now be transformed directly to the
Fourier-mode basis.

\subsubsection{Mode algebra in the forward representation}
\label{app:Vir-forward}

Applying the Fourier expansion introduced in
Appendix~\ref{app:KM-Fourier} to
Eq.~\eqref{eq:app-Vir-DS}, the three contributions to the mode
transformation can be evaluated separately.

For the first term, using
Eq.~\eqref{eq:app-Vir-Fourier-L}, one obtains
\begin{align}
\frac{1}{N}\sum_{n=0}^{N-1}
e^{-ir\tau_n}\,
\epsilon_n^{(s)}
\nabla\mathcal L_n
&=
-\frac{i}{N}
\sum_{n,p}
e^{-ir\tau_n}
e^{-is\tau_n}
\omega_p^{(+)}
L_p e^{ip\tau_n}
\nonumber\\
&=
-i\sum_p
\omega_p^{(+)}
L_p\,
\delta_{r+s-p,0}
\nonumber\\
&=
-i\omega_{r+s}^{(+)}
L_{r+s}.
\label{eq:app-Vir-forward-first}
\end{align}

The second term gives
\begin{align}
\frac{2}{N}\sum_{n=0}^{N-1}
e^{-ir\tau_n}\,
\mathcal L_n
\nabla\epsilon_n^{(s)}
&=
-\frac{2i}{N}
\sum_{n,p}
e^{-ir\tau_n}
L_p e^{ip\tau_n}
\omega_{-s}^{(+)}
e^{-is\tau_n}
\nonumber\\
&=
-2i\omega_{-s}^{(+)}
\sum_p
L_p\,
\delta_{r+s-p,0}
\nonumber\\
&=
-2i\omega_{-s}^{(+)}
L_{r+s}.
\label{eq:app-Vir-forward-second}
\end{align}

For the last term,
Eq.~\eqref{eq:app-Vir-forward-action} yields
\begin{align}
\frac{2}{\alpha N}
\sum_{n=0}^{N-1}
e^{-ir\tau_n}
\nabla^3\epsilon_n^{(s)}
&=
-\frac{2i}{\alpha N}
\left(\omega_{-s}^{(+)}\right)^3
\sum_{n=0}^{N-1}
e^{-i(r+s)\tau_n}
\nonumber\\
&=
-\frac{2i}{\alpha}
\left(\omega_{-s}^{(+)}\right)^3
\delta_{r+s,0}.
\label{eq:app-Vir-forward-third}
\end{align}

Combining
Eqs.~\eqref{eq:app-Vir-forward-first}--\eqref{eq:app-Vir-forward-third},
the Fourier-mode form of the lattice Virasoro transformation in the forward
representation is
\begin{equation}
\delta_s L_r
=
-i\left[
\omega_{r+s}^{(+)}
+
2\omega_{-s}^{(+)}
\right]
L_{r+s}
-
i\frac{2}{\alpha}
\left(\omega_{-s}^{(+)}\right)^3
\delta_{r+s,0}.
\label{eq:app-Vir-forward-mode}
\end{equation}

The continuum limit of
Eq.~\eqref{eq:app-Vir-forward-mode}
is considered in the following subsection.
\subsubsection{Continuum limit of the forward representation}
\label{app:Vir-forward-continuum}

For the forward lattice momentum,
\begin{equation}
\omega_q^{(+)}
=
iq+\mathcal O(\Delta\tau),
\qquad
\Delta\tau\rightarrow0,
\label{eq:app-Vir-forward-omega-limit}
\end{equation}
and therefore
\begin{align}
\omega_{r+s}^{(+)}
+
2\omega_{-s}^{(+)}
&=
i(r+s)-2is+\mathcal O(\Delta\tau)
\nonumber\\
&=
i(r-s)+\mathcal O(\Delta\tau).
\label{eq:app-Vir-forward-structure-limit}
\end{align}
Similarly,
\begin{equation}
\left(\omega_{-s}^{(+)}\right)^3
=
is^3+\mathcal O(\Delta\tau).
\label{eq:app-Vir-forward-central-limit}
\end{equation}
Hence Eq.~\eqref{eq:app-Vir-forward-mode} becomes
\begin{equation}
\left\{L_r,L_s\right\}
=
(r-s)L_{r+s}
+
\frac{2}{\alpha}
s^3\delta_{r+s,0}
+
\mathcal O(\Delta\tau).
\label{eq:app-Vir-forward-continuum}
\end{equation}
The continuum transformation thus exhibits the characteristic Virasoro
mode structure $(r-s)L_{r+s}$, with the central contribution proportional
to the cubic mode number.
Using $s=-r$ on the support of the Kronecker delta, the central term may
equivalently be written as
\begin{equation}
\left\{L_r,L_s\right\}
=
(r-s)L_{r+s}
-
\frac{2}{\alpha}
r^3\delta_{r+s,0}
+
\mathcal O(\Delta\tau).
\label{eq:app-Vir-forward-continuum-r}
\end{equation}
A shift of the zero mode,
\begin{equation}
L_r
\;\longrightarrow\;
L_r-\frac{1}{\alpha}\delta_{r,0},
\label{eq:app-Vir-zero-mode-shift}
\end{equation}
then brings the algebra to the standard form
\begin{equation}
\left\{L_r,L_s\right\}
=
(r-s)L_{r+s}
-
\frac{2}{\alpha}
\left(r^3-r\right)
\delta_{r+s,0}
+
\mathcal O(\Delta\tau).
\label{eq:app-Vir-standard}
\end{equation}
Comparing the central term with the standard Virasoro normalization,
\begin{equation}
\frac{c}{12}\left(r^3-r\right)\delta_{r+s,0},
\end{equation}
fixes
\begin{equation}
\alpha=-\frac{24}{c}.
\end{equation}
Using the standard normalization $c=6k$, this gives
\begin{equation}
\alpha=-\frac{4}{k}.
\end{equation}

\subsection{Comparison with Previous Lattice Constructions}
\label{app:previous-lattice}

Earlier lattice realizations of current and Virasoro algebras were developed
from rather different starting points. Two representative constructions are
those of Faddeev and Volkov~\cite{Faddeev:1993pe} and Koo and
Saleur~\cite{Koo:1993wz}. Although both recover continuum conformal
structures, their lattice variables and the route to the continuum limit
differ from those used in the present BF formulation.

Faddeev and Volkov~\cite{Faddeev:1993pe} construct a lattice analogue of an
Abelian current algebra and show that, together with a discrete-time
evolution, it gives rise to a lattice Virasoro algebra and an associated
hierarchy of conserved quantities. The Virasoro structure is therefore built
directly from a lattice current algebra within an integrable lattice
framework.

Koo and Saleur~\cite{Koo:1993wz}, on the other hand, study the emergence of
Virasoro symmetry in the scaling limit of XXZ- and RSOS-type lattice models.
Their construction is based on lattice approximants to the stress-energy
tensor, whose algebraic relations reproduce the Virasoro algebra in the
continuum scaling limit. In this approach the underlying lattice degrees of
freedom are those of the corresponding statistical or quantum spin-chain
models.

The present construction starts instead from a discrete BF gauge theory. The
basic variables are lattice holonomies together with the adjoint dilaton
field, and the boundary current algebra follows from the residual lattice
gauge transformations. Passing to Fourier modes gives the lattice affine
Kac--Moody structure discussed in Appendix~\ref{app:lattice-KM}. The
Drinfeld--Sokolov reduction then leads to the lattice Virasoro transformation
analyzed in Appendix~\ref{app:lattice-Virasoro}, with the conventional affine
and Virasoro structures recovered in the continuum limit.

Thus, the three constructions share the appearance of Virasoro symmetry from
a finite lattice description, but differ in their fundamental lattice
variables and in the mechanism by which the continuum conformal algebra is
reached. In the present case, the construction is tied directly to the
holonomy and gauge structure of the discrete BF theory.

\color{black}

\end{document}